\documentclass[sigconf]{acmart}
\AtBeginDocument{%
  }

\setcopyright{acmlicensed}
\copyrightyear{2026}
\acmYear{2026}
\acmDOI{XXXXXXX.XXXXXXX}
\acmConference[SIGMOD '26]{International Conference on Management of Data}{May 31--June 05,
	2026}{Bengaluru, India}
\acmISBN{978-1-4503-XXXX-X/2026/05}
\usepackage{tikz}
\usepackage{amsmath}
\usepackage{xspace}
\usepackage{enumitem}

\usepackage{filecontents}

\usepackage{pifont}
\usepackage{txfonts}
\usepackage{array, boldline, makecell}
\usepackage{circledsteps}
\usepackage{color}
\usepackage{booktabs}
\usepackage{caption}
\usepackage{multirow}
\usepackage{subfig}
\usepackage{url}
\usepackage{textcomp}
\usepackage{tabularx}
\usepackage{txfonts}
\usepackage{amssymb}
\usepackage{amsthm}

\usepackage{amsmath}
\usepackage{graphicx}
\usepackage{xcolor}

\newcommand{\sys}{\textit{Beluga}\xspace}

\usepackage{color}
\definecolor{rcolor}{RGB}{20,99,255} 

\definecolor{color1}{RGB}{0,114,178}    % 深蓝
\definecolor{color2}{RGB}{0,158,115}    % 深绿
\definecolor{color3}{RGB}{255,165,0}     % 深橙
\definecolor{color4}{RGB}{128,0,0}  % 暗红
\newcommand{\revisiona}[1]{{#1}}
\newcommand{\revisionb}[1]{{#1}}
\newcommand{\revisionc}[1]{{#1}}
\newcommand{\revisionx}[1]{{#1}}

\newcommand{\mycaption}[3]{{\vspace{-0.1cm} \caption{\label{#1}{#2. }{ #3}} \vspace{-0.15cm}}}
\newcommand{\tblcaption}[3]{{\caption{\label{#1}{#2. }{ #3}} \vspace{-0.3cm} } }

\usepackage{tikz,lipsum}
\usepackage[most]{tcolorbox}
\usepackage{mdframed}
\newtcolorbox{myframe}[2][]{%
	colback=gray!10,%gray background white
	enhanced,colframe=black,coltitle=black,
	sharp corners,boxrule=0.4pt,
	fonttitle=\bfseries,
	attach boxed title to top left={yshift=-0.3\baselineskip-0.4pt,xshift=2mm},
	boxed title style={tile,size=tight,width=fit to content,left=0.5mm,right=0.5mm,
		colback=white,before upper=\strut},
	title={\parbox{0.95\textwidth}{#2}},#1
    before upper={\setlength{\parindent}{1em}}  % 使用文档默认缩进值
}
\usepackage{threeparttable}
\tcbuselibrary{listings,breakable}
\usepackage{tcolorbox}
\usepackage{lipsum}

\usepackage[linesnumbered,ruled,vlined]{algorithm2e}

\begin{document}
% \input{sec/revision.tex}
%%%%%%%%%%%%%%%%%%%%%%%%%%%%%%%%%%%%%%%%%%%%%%%%%%%%%%%%%%%%%%%%%%%%%%%%%%%%%%%%

%%%%%%%%%%%%%%%%%%%%%%%%%%%%%%%%%%%%%%%%%%%%%%%%%%%%%%%%%%%%%%%%%%%%%%%%%%%%%%%%
% \begin{@twocolumnfalse}
%   \vspace{0.3in}
    % \section{Response to Revision Instructions}
    % {\centering \part*{ Response to Revision Instructions}}
    % \label{appendix}
% We sincerely thank all reviewers for their valuable comments and suggestions, which have helped us significantly improve our paper.
% Important modifications in the revised manuscript are highlighted in yellow.
% We thank the reviewers and meta-reviewers for the valuable feedback and appreciate the great efforts in reviewing our paper. 
% We have prepared a revised version to address the questions raised, with the necessary modifications highlighted in \textcolor{blue}{blue}. 
% % We address reviewers' comments in detail as below:
% \vspace{0.1in}
% \hrule
% \hrule
% \vspace{0.2in}
% % \end{@twocolumnfalse}
% % ]
% \input{sec/revision.tex}

%don't want date printed
\date{}

% make title bold and 14 pt font (Latex default is non-bold, 16 pt)
\title{\sys: A CXL-Based Memory Architecture for Scalable and Efficient LLM KVCache Management}

%for single author (just remove % characters)
% \author{
% {\rm Submission \#70 (Major Revision of FAST'25 \#111)}
% \thanks{\autoref{appendix} Response to Revision Instructions}}

% \author{PolarDB, Alibaba Cloud Computing} 

% \author{\rm Xinjun Yang, Qingda Hu, Junru Li, Feifei Li, Yuqi Zhou, Yicong Zhu, Qiuru Lin, Jian Dai, Yang Kong, Jiayu Zhang, Guoqiang Xu, Qiang Liu}
% % \email{trovato@corporation.com}
% % \orcid{1234-5678-9012}
% % \author{G.K.M. Tobin}
% % \authornotemark[1]
% % \email{webmaster@marysville-ohio.com}
% \affiliation{%
% % \large Alibaba Cloud Computing
%   \institution{Alibaba Cloud Computing}
%   % \city{Dublin}
%   % \state{Ohio}
%   % \country{USA}
% }

\author{Xinjun Yang}
\affiliation{%
  \institution{Alibaba Cloud Computing}
  \city{Sunnyvale}
  \state{CA}
  \country{USA}
}
\author{Qingda Hu}
% \authornote{Qingda Hu is the corresponding author (rusuo.ljr@alibaba-inc.com).}
\affiliation{%
  \institution{Alibaba Cloud Computing}
  \city{Hangzhou}
  \country{China}
}
\authornote{Qingda Hu and Junru Li are the corresponding authors \\ \ \ \{qingda.hqd, rusuo.ljr\}@alibaba-inc.com}
\author{Junru Li}
\authornotemark[1]
% \authornote{Junru Li is the corresponding author (rusuo.ljr@alibaba-inc.com).}
\affiliation{%
  \institution{Alibaba Cloud Computing}
  \city{Beijing}
  \country{China}
}
\author{Feifei Li}
\affiliation{%
  \institution{Alibaba Cloud Computing}
  \city{Hangzhou}
  \country{China}
}
\author{Yicong Zhu}
\affiliation{%
  \institution{Alibaba Cloud Computing}
  \city{Hangzhou}
  \country{China}
}
\author{Yuqi Zhou}
\affiliation{%
  \institution{Alibaba Cloud Computing}
  \city{Beijing}
  \country{China}
}
\author{Qiuru Lin}
\affiliation{%
  \institution{Alibaba Cloud Computing}
  \city{Hangzhou}
  \country{China}
}
\author{Jian Dai}
\affiliation{%
  \institution{Alibaba Cloud Computing}
  \city{Hangzhou}
  \country{China}
}
\author{Yang Kong}
\affiliation{%
  \institution{Alibaba Cloud Computing}
  \city{Shanghai}
  \country{China}
}
\author{Jiayu Zhang}
\affiliation{%
  \institution{Alibaba Cloud Computing}
  \city{Shanghai}
  \country{China}
}
\author{Guoqiang Xu}
\affiliation{%
  \institution{Alibaba Cloud Computing}
  \city{Hangzhou}
  \country{China}
}
\author{Qiang Liu}
\affiliation{%
  \institution{Alibaba Cloud Computing}
  \city{Shenzhen}
  \country{China}
}
\renewcommand{\shortauthors}{Xinjun Yang, Qingda Hu, Junru Li, Feifei Li et al.}


\settopmatter{authorsperrow=4}

\setcounter{page}{1}

%-------------------------------------------------------------------------------
\begin{abstract}
The rapid increase in LLM model sizes and the growing demand for long-context inference have made memory a critical bottleneck in GPU-accelerated LLM serving. Although high-bandwidth memory (HBM) on GPUs offers fast access, its limited capacity necessitates reliance on host memory (CPU DRAM) to support large KVCache.
% While high-bandwidth memory (HBM) on GPUs offers fast access, its capacity is limited—typically up to 192GB—necessitating reliance on host memory (CPU DRAM) to support larger working sets such as the KVCache. 
However, the maximum DRAM capacity is constrained by the limited number of memory channels per CPU socket. To overcome this limitation, current systems often adopt RDMA-based disaggregated memory pools, which introduce significant challenges including high access latency, complex communication protocols, and synchronization overhead. Fortunately, the emerging CXL technology introduces new opportunities in KVCache design. In this paper, we propose \sys, a novel memory architecture that enables GPUs and CPUs to access a shared, large-scale memory pool through CXL switches. By supporting native load/store access semantics over the CXL fabric, our design delivers near-local memory latency, while reducing programming complexity and minimizing synchronization overhead. We conduct a systematic characterization of CXL-based memory pool and propose a set of design guidelines. Based on \sys, we design and implement \sys-KVCache, a system tailored for managing the large-scale KVCache for LLM inference. \sys-KVCache achieves an 89.6\% reduction in Time-To-First-Token (TTFT) and 7.35$\times$ throughput improvement in vLLM compared to RDMA-based solutions. To the best of our knowledge, \sys is the first system that enables GPUs to directly access large-scale memory pools through CXL switches, marking a significant step toward low-latency, shared access to vast memory resources by GPUs.

\end{abstract}
\maketitle

\section{Introduction}
\begin{figure}[t]
  \centering
  \subfloat[RDMA-based MemPool]{\includegraphics[trim=0 0 14.4cm 0,clip,width=0.5\linewidth]{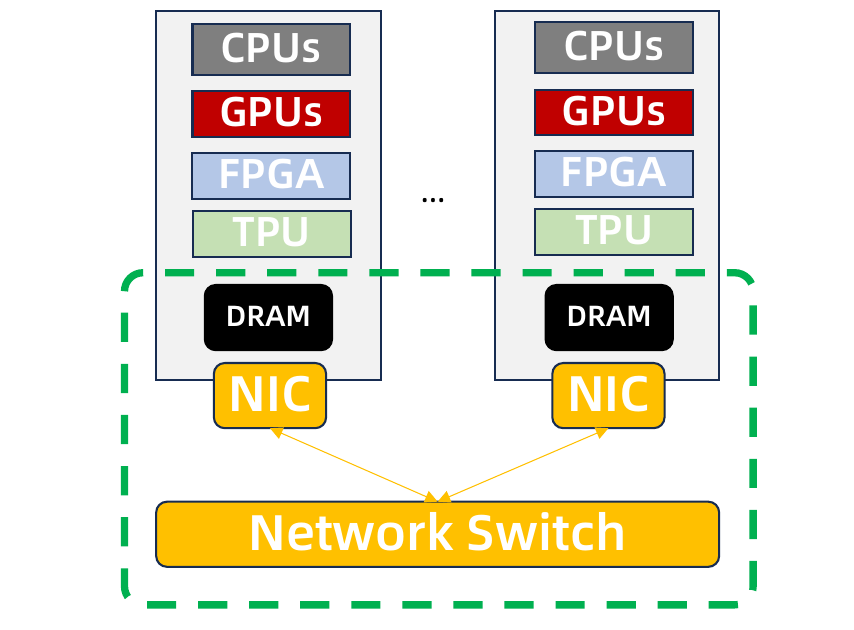}\label{fig1:rdmaM}}
  \hfill
  \subfloat[\sys]{\includegraphics[trim=14.47cm 0 0in 0,clip,width=0.5\linewidth]{pic/mempool.pdf}\label{fig1:cxlM}}
  \vspace{-0.2cm}
  \mycaption{pic:mempool}{Overview of RDMA/CXL memory pools}{}
  \vspace{-0.1cm}
\end{figure}

Database vendors are rapidly integrating in-database (in-DB) LLM inference capabilities~\cite{DBLP:journals/pacmmod/GiannakourisT25,10.1145/3725262,10.1145/3725356,DBLP:journals/pvldb/ZhouCDMYZZ22,DBLP:journals/pvldb/SalazarDiazGR24}, as demonstrated by MindsDB~\cite{mindsdb}, Aurora~\cite{Aurora}, PolarDB~\cite{PolarDB}, and GaussDB~\cite{GaussDB}. These systems leverage GPU-CPU platforms to enable advanced applications, particularly Retrieval-Augmented Generation (RAG) based intelligent Q\&A~\cite{DBLP:conf/vldb/NieHSWJZZS24,DBLP:journals/pvldb/ZhangHFCDP24} and natural language to SQL translation (NL2SQL)~\cite{10.1145/3725331,DBLP:conf/iclr/LongGSYWW0025}, enabling interactive and intelligent data exploration. These applications impose significant demands on the underlying systems to efficiently process long and complex contexts. A critical optimization is reusing the model’s internal representation of context—namely, the Key-Value (KV) Cache—which avoids redundant and costly recomputation of shared contexts~\cite{DBLP:journals/corr/abs-2407-18003,DBLP:conf/mlsys/AdnanAJNSK24}. However, KVCache maintenance demands significant memory resources~\cite{DBLP:conf/nips/HooperKMMSKG24,cachegen}. 
For instance, 50M tokens in Kimi~\cite{kimi,mooncake} requires approximately 20TB of DRAM for its KVCache to achieve maximal cache hit ratios. Given ever-increasing inference memory demands, local GPU clusters are becoming financially and technically infeasible to meet the rising requirements of LLM inference. Consequently, remote resource pooling architectures centered around KVCache~\cite{DBLP:journals/corr/abs-2501-14743}, such as Dynamo~\cite{Dynamo} and MoonCake~\cite{mooncake,mooncake}, are gaining considerable traction and real-world adoption. By offloading the KVCache to high-capacity remote memory pools via Remote Direct Memory Access (RDMA)~\cite{Dynamo, mooncake,mooncake, DBLP:journals/corr/abs-2501-14743}, these architectures achieve enhanced memory scalability and system efficiency, as shown in Figure~\ref{fig1:rdmaM}.

% Large Language Models (LLMs) are now widely used in a variety of applications.  
% Among them, some applications like multi-turn conversations and Retrieval-Augmented Generation (RAG) require LLMs to process shared contexts for different requests, e.g., the dialogue history in multi-turn conversations or the retrieved documents in RAG.
% To avoid redundant and costly re-computation of the shared context, modern inference systems rely on a critical optimization: reusing the model's internal representation, known as the Key-Value (KV) Cache~\cite{vllm,DBLP:conf/nips/ZhengYXS0YCKSGB24}. It is essential for accelerating subsequent inference request with the same prefix.  
% However, the size of the KVCache scales linearly with the sequence length of the prompt. For long-context applications, this can lead to a memory footprint that far exceeds the capacity of a single server's GPU HBM and host memory. For instance, recent work from Mooncake~\cite{mooncake} highlights that supporting the multi-turn conversations in Kimi~\cite{} requires approximately 20TB of storage for its KVCache to achieve the maximum cache hit ratio.

% To mitigate this memory bottleneck, a common strategy is to offload the KVCache to a remote memory pool using Remote Direct Memory Access (RDMA)~\cite{Dynamo, mooncake, DBLP:journals/corr/abs-2501-14743}, as illustrated in Figure~\ref{fig:rdmaM}. 
However, RDMA was fundamentally designed as a networking protocol, not a memory bus~\cite{DBLP:conf/osdi/WeiD0C18,DBLP:conf/osdi/WeiLW0Y0023, StreamRDMA}, leading to some limitations for KVCache offloading. On the data path, it requires extra data movement through host memory, which in turn incurs extra latency. On the control path, managing RDMA communication involves complex programming, introduces overhead for preparing requests and polling for completions, and necessitates costly cross-component synchronization. 
Furthermore, the non-uniform hierarchy between local and remote memory creates a challenging scheduling problem, requiring a balance between computing resources and KVCache locality~\cite{mooncake,mooncake, Dynamo}. 
These combined inefficiencies in RDMA memory pool significantly limit the potential benefits of KVCache offloading. 
%These combined inefficiencies in RDMA memory pool overshadow the potential benefits of KVCache offloading. 

The emergence of the Compute Express Link (CXL) standard presents an opportunity to design high-performance memory pools. By providing a direct, low-latency, load/store memory interface, CXL allows CPU/GPUs to access remote memory with an efficiency that approaches local memory access, as shown in Figure~\ref{fig1:cxlM}. 
This memory-semantic interface directly addresses the core problems of RDMA: it simplifies the data path by eliminating extra copies, and streamlines the control path by removing the need for complex network programming and explicit synchronization.

While the potential of CXL is widely discussed, prior work has focused on memory expansion with CXL 1.1 devices or FPGA-based simulations. A practical, large-scale evaluation of CXL 2.0 memory pools with multiple hosts and devices has not been possible due to the lack of commercial hardware.
The recent availability of the XConn XC50256 CXL 2.0 switch~\cite{xconn} now enables such a study. This paper presents \sys, a system built on this new CXL 2.0 switch hardware. We first provide a detailed characterization of \sys, measuring its performance with both CPU and GPU workloads to understand its fundamental behavior and propose design guidelines. Building on these findings, we then implement \sys-KVCache and integrate it into the vLLM inference engine. This demonstrates the effectiveness of our proposed optimizations for KVCache management, a critical component in LLM serving.
% The potential of CXL has been widely discussed, prior work~\cite{cxlrack, cxlpci,DBLP:conf/asplos/LiuHWBNJNL25, pond, gouk2022direct} has focused on memory expansion with CXL 1.1 devices or FPGA-based simulations of shared CXL devices. However, a practical, large-scale evaluation of CXL 2.0 memory pools (supporting multiple CXL devices and multiple hosts) has remained infeasible due to the lack of production-ready hardware.
% The recent availability of XConn B1 CXL 2.0 switches now enables such an empirical study~\cite{xconn, polarcxl}. 
% This paper presents, 
% \sys, a novel memory architecture, that leverages XConn B1 CXL 2.0 switches to provide a shared memory pool for KVCache Management. 
% We first present an in-depth systematic characterization of \sys. We evaluate its performance with both CPU and GPU workloads to understand its fundamental characteristics. Building on these insights, we integrate \sys into the vLLM inference engine to analyze its efficacy for KVCache offloading and reuse. 

\noindent
   We summarize our main contributions as follows:
\begin{itemize}
[itemsep=0pt, parsep=0pt, labelsep=3pt, leftmargin=*, topsep=0pt, partopsep=0pt]
    \item We present \sys, a memory system for GPU clusters that leverages Compute Express Link (CXL) 2.0 switches to provide access to a disaggregated, shared memory pool. To the best of our knowledge, \sys is the first system to integrate GPU clusters with CXL switching infrastructure, demonstrating the feasibility of extending GPU cluster memory capacity through CXL-based interconnects. 
    
    \item We conduct comprehensive performance evaluations of \sys on GPU clusters accessing memory via a CXL 2.0 switch, and we introduce a set of system-level optimizations and guidelines to address key performance bottlenecks. These optimizations enable \sys to effectively overcome the cache coherence limitations inherent in a multi-host CXL 2.0 environment. Compared to RDMA-based disaggregated memory systems, \sys achieves up to 7.0$\times$ lower latency for write operations and up to 6.3$\times$ lower latency for read operations. Additionally, \sys delivers scalable memory bandwidth, making it suitable for high-performance, data-intensive GPU workloads.    

    \item We propose \sys-KVCache, a KVCache management system built on the CXL memory pool. We demonstrate its effectiveness by integrating it into vLLM~\cite{vllm}, the open-source inference library, delivering efficient KVCache read/write, CXL-based RPCs, and a simplified scheduler. Our evaluation shows that \sys-KVCache improves LLM inference throughput by up to 4.79$\times$ over MoonCake, the state-of-the-art RDMA-based solution.
\end{itemize}

\noindent

This paper is organized as follows.
Section~\ref{sec:bg} presents the background of this paper.
\revisionc{Section \ref{gpunic} reviews RDMA-based memory pooling and presents our motivation, then Section \ref{gpucxl} introduces the architectural design of our system \sys and its inherent advantages.}
Section~\ref{sec:featofsys} details its characterization and optimizations, while Section~\ref{sec:kvcache} presents our KVCache management system. 
Section~\ref{sec:evaluation} then presents an end-to-end evaluation. Section~\ref{sec:future} discusses the broader implications of our work, and Section~\ref{sec:relatedwork} reviews the related work.

\section{Background}
\label{sec:bg}
\subsection{LLM Inference and KVCache}
Large Language Models (LLMs), powered by the Transformer architecture, have become a foundational workload for modern applications~\cite{yang2025qwen3,achiam2023gpt,dubey2024llama}. Their inference process typically consists of two phases: prefill and decode~\cite{DBLP:conf/osdi/ZhongLCHZL0024}. The compute-intensive prefill phase processes the input prompt in parallel to generate the initial output token. Subsequently, the decode phase generates the output sequence autoregressively, where each new token's generation depends on all preceding tokens and their corresponding KV activations.

The KVCache, which stores intermediate KV activations from previous tokens, serves as an essential space-for-time optimization~\cite{vllm,DBLP:conf/nips/ZhengYXS0YCKSGB24}. 
Eliminating redundant computation is essential for achieving low-latency token generation. This technique is widely used in various scenarios: preserving history in multi-turn dialogues, enabling inter-request sharing for fixed contexts (e.g., system prompts), and amortizing the cost of processing long documents in RAG~\cite{gao2023retrieval}.
These scenarios reveal a new direction of data management: designing efficient storage systems to offload, reuse, and share KVCache~\cite{cachegen,mooncake,mooncake,yao2025cacheblend,DBLP:conf/osdi/LeeLSS24,hu2024memserve}. Such systems have the following four fundamental requirements:

\noindent
\textbf{Scalable capacity.} The memory footprint of the KVCache increases linearly with context length, frequently consuming gigabytes of memory per request. The disaggregated system must therefore provide a large, elastic memory pool capable of scaling beyond the limitations of a single node’s on-package memory.

\noindent
\textbf{Efficient sharing.} Optimizing resource utilization and throughput in multi-server GPU clusters necessitates a disaggregated architecture, in which all servers access a unified KVCache memory pool.

\noindent
\textbf{Low-latency access.} The primary benefit of employing a cached prefix is to eliminate the overhead of recomputation, which directly enhances metrics like Time-To-First-Token (TTFT). Consequently, the end-to-end latency for KVCache retrieval from the memory pool must be substantially lower than its recomputation latency. 

\noindent
\textbf{High aggregate throughput.} The inherent parallelism within a multi-GPU server necessitates a high-throughput requirement for the memory system. Insufficient aggregate bandwidth will lead to GPU stalling, directly constraining system throughput and potentially rendering recomputation a more efficient alternative. 

\begin{figure*}[t!]
  \centering
  \subfloat[GPU clusters with an RDMA-based memory pool]{\includegraphics[trim=0 0 17.5cm 0,clip,width=0.45\linewidth]{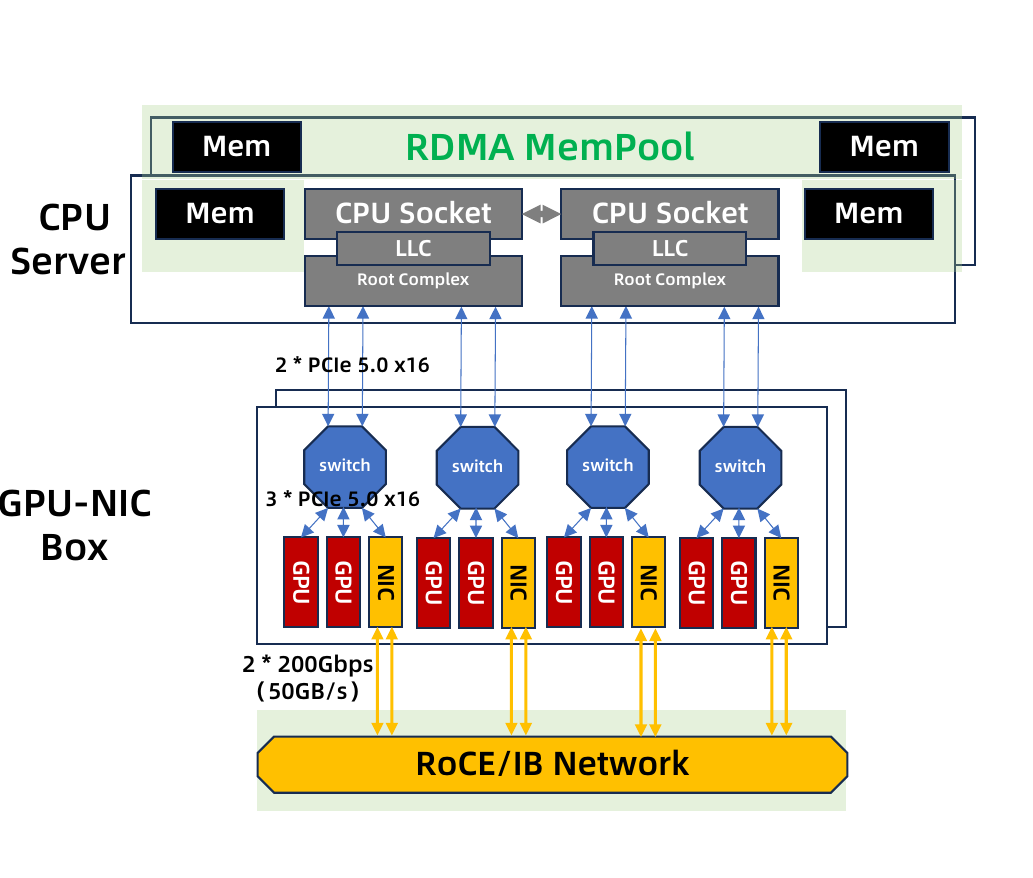}\label{fig:rdmaM}}
  \hfill
  \subfloat[GPU clusters with a CXL-based memory pool (\sys)]{\includegraphics[trim=17.5cm 0 0in 0,clip,width=0.45\linewidth]{pic/cxl.pdf}\label{fig:cxlM}}
  \vspace{-0.1cm}
  \mycaption{pic:arch}{Hardware architectures of GPU clusters with (a) RDMA / (b) CXL memory pool}{The real hardware is shown in~\autoref{pic:arch}.} 
\end{figure*}

% \noindent
\subsection{Remote Direct Memory Access (RDMA)}
% \textbf{RDMA.}
RDMA enables a Network Interface Controller (NIC) to directly access memory on a remote host, bypassing the remote CPU and the OS kernel~\cite{rdmaMpi,rdmadb,bin1, bin2, ziegler2023design}. By minimizing network stack overhead, RDMA achieves high bandwidth and low CPU utilization, making it a widely adopted interconnect in modern datacenters. Most RDMA memory pools rely on one-sided communication primitives, which facilitate direct memory-to-memory data transfers between nodes. This architectural design eliminates traditional overheads, such as kernel context switches and data copies. Key primitives include RDMA Read and RDMA Write for direct data retrieval and placement, and RDMA Atomic Operations (e.g., CAS, FAA) for executing atomic read-modify-write on remote memory. 
% These atomics are crucial for implementing efficient distributed synchronization mechanisms like locks. 
RDMA's one-sided primitives provide an ultra-low latency, high-throughput interconnect, forming the foundation for building high-performance memory pools.

\subsection{Compute Express Link (CXL)}

% \noindent
% \textbf{CXL.}
CXL is an emerging interconnect protocol designed to facilitate efficient memory sharing between host CPUs and accelerators. It defines three distinct protocols: CXL.IO, dedicated to CXL device management and DMA data transfers; CXL.cache, for coherent device access to host memory; and CXL.mem, enabling host load/store access to device-attached memory for expansion and pooling. The standard's rapid evolution has progressively enhanced system capabilities: transitioning from single-device attachment (CXL 1.1), to switch-based memory pooling (CXL 2.0), and most recently to multi-tier switching with cache coherency support (CXL 3.0), thereby enabling more flexible and powerful heterogeneous systems.

This paper focuses primarily on memory expansion solutions leveraging the CXL.mem interface. Currently, the design and production of hardware supporting CXL.mem remains in its early stages. Most existing hardware only supports CXL 1.1, providing memory expansion capabilities for single servers. Although several research efforts have developed FPGA-based prototypes supporting CXL 2.0, their capabilities are constrained in both device count and memory capacity~\cite{cxlrack, cxlpci,DBLP:conf/asplos/LiuHWBNJNL25, pond, gouk2022direct}. XConn~\cite{xconn,polarcxl} has produced the first commercial CXL 2.0 switch, providing CXL.mem interfaces with support for \textit{256 lanes} and concurrent access from multiple hosts, delivering minimal 64-byte I/O latency of \textasciitilde \textit{750ns} and maximum switching capacity of \textit{2TB/s}.
The introduction of CXL 2.0 switches fundamentally elevates the protocol beyond the single-node memory expansion of CXL 1.1, thereby enabling the creation of a large-scale shared memory pool.

%Compute Express Link (CXL) is an emerging interconnect protocol that creates new opportunities for system design. The CXL protocol comprises three types: CXL.IO, CXL.Cache, and CXL.Mem. CXL.IO, similar to traditional PCIe, is used for CXL device management and DMA data transfers. CXL.Cache enables device access to host cache, facilitating high-performance storage/network/compute hardware drivers and data transfer mechanisms. CXL.Mem provides efficient load/store access interfaces for host memory expansion hardware. The CXL standard has evolved significantly: from single-host CXL devices (CXL 1.1, 2019) to switch-based memory pooling (CXL 2.0, 2020), and further to multi-tier switching with cache coherency support (CXL 3.0, 2022).

% Our work is the first to harness this new capability with commercial CXL 2.0 hardware to design and evaluate a shared memory architecture specifically for accelerating LLM inference.

% as shown in Figure \ref{fig:mempconsumption}.

% The figure needs to be redrawn 
% \begin{figure}[]
%   \begin{center}
%   \includegraphics[width=\linewidth]{pic/memconsumption.png}
%   \end{center}
%   \mycaption{fig:mempconsumption}{Memory Consumption of LLM}{} 
% \end{figure}

% \section{\sys: A CXL-Based Memory Architecture}
%\section{Motivation and Design}
%\section{Overcoming RDMA Limitations with CXL}
%\section{\sys: A CXL-Based Memory Architecture}
%\section{The CXL MemPool for in-DB LLM Inference}
%\section{Motivation of CXL MemPool}
\label{sec:gpu-cxl-arch}

% The conventional approach, shown in Figure~\ref{fig:rdmaM}, uses an RDMA-based memory pool to aggregate DRAM across the cluster for storing the offloaded KVCache.
% The RDMA architecture offers two main benefits: scalable memory capacity and data sharing across servers. 
% However, this design suffers from fundamental limitations, including performance bottlenecks from extra data copies, expensive control-path synchronization and high architectural complexity (\S\ref{gpunic}).

% To overcome these challenges, we propose \sys, a CXL-based memory pool that uses a CXL switch to create a scalable, efficient memory space (Figure~\ref{fig:cxlM}).
% This allows GPUs to access pooled memory with simple load/store operations, solving the performance bottlenecks and programming complexity of the RDMA approach (\S\ref{gpucxl}).

\section{\revisionc{RDMA-Based MemPool}}
\label{gpunic}

The conventional approach for KVCache management, shown in Figure~\ref{fig:rdmaM}, uses an RDMA-based memory pool to aggregate DRAM across the cluster for storing the offloaded KVCache.
The RDMA architecture offers two main benefits: scalable memory capacity and data sharing across servers. 
However, this design suffers from fundamental limitations, including performance bottlenecks from extra data copies, expensive control-path synchronization and high architectural complexity (\S\ref{gpunic}).

\subsection{A Simple Review of RDMA-Based MemPool}

Figure~\ref{fig:rdmaM} shows a multi-GPU cluster interconnected by an RDMA network. This architecture is common in modern data centers~\cite{liu2024deepseek}.
Each server in the cluster contains a host CPU, multiple GPUs, and RDMA NICs. These components are interconnected by a fabric of PCIe switches. 
A typical 8-GPU server, for example, uses four such switches. Each switch provides five PCIe x16 links. Three links connect to two GPUs and an RDMA NIC. The remaining two serve as uplinks to a host CPU socket.
%Each switch connects to two GPUs and one RDMA NIC using three PCIe x16 links, and also connects to a CPU socket with two additional PCIe x16 links.
The dedicated RDMA NICs on each server provide up to 1.6 Tbps of inter-server bandwidth. This high-speed interconnect allows each server to contribute local DRAM (e.g., 2 TB), forming a distributed memory pool shared across the entire cluster.
To leverage this shared memory pool, LLM inference frameworks access the distributed memory pool using two main approaches: a CPU-driven approach or a GPU-driven one. The key distinction is which processor orchestrates the data transfers.

\begin{figure}[t!]
  \centering
    \includegraphics[width=0.9\linewidth]{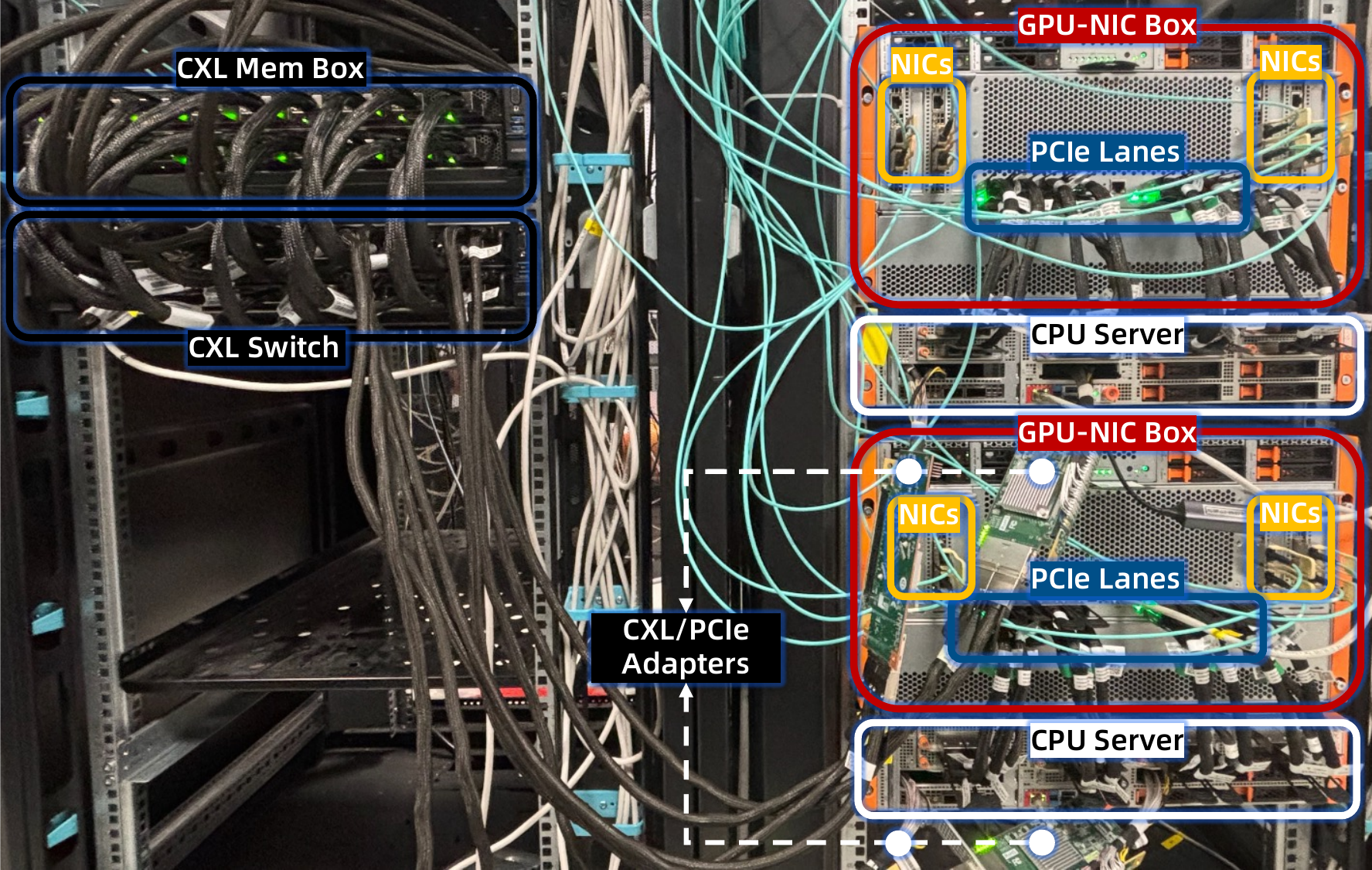}
  \vspace{-0.1cm}
  \mycaption{pic:arch}{The actual deployment of \sys}{} 
\end{figure}

\noindent\textbf{CPU-driven RDMA memory pool access.}
% Figure~\ref{fig:a} illustrates the most common strategy for KVCache offloading: a CPU-driven model. 
This approach is adopted by prominent frameworks such as vLLM~\cite{vllm}, MoonCake~\cite{mooncake,mooncake}, and LMCache~\cite{cachegen}. 
In this model, the host CPU issues RDMA commands directly to the NIC to fetch remote data.
The process of storing the KVCache involves two steps. First, data is copied from the GPU to a "bounce buffer" in host memory. The CPU then initiates an RDMA transfer to move data from this buffer to the remote pool. Loading data follows the reverse path: from the remote pool, through the bounce buffer, and finally back to the GPU.
%In this design, all data is staged through the CPU's main memory. To store the KV Cache, data travels from the GPU to a "bounce buffer" in host memory, after which the CPU initiates an RDMA transfer to the remote pool.  The loading process simply reverses this path, bringing data back through the same bounce buffer to the GPU. 

\noindent\textbf{GPU-driven RDMA memory pool access.}
Alternatively, a GPU-driven approach uses GPUDirect RDMA (GDR)~\cite{gdr}.
% ,as shown in Figure~\ref{fig:b}. 
This technology allows the GPU to issue commands directly to the NIC, completely bypassing the host CPU.
In this model, a dedicated GPU kernel manages all network communication. To store the KVCache, this kernel issues an RDMA Write and polls for its completion. To read the cache, it issues an RDMA Read and polls for a work completion (WC) signal. Once the data arrives, the polling kernel must synchronize with the main computation stream to proceed with the transformer computation. 
Note that GDR is a feature exclusive to expensive datacenter-class GPUs like the NVIDIA A100/H100. It is not available on widely-used consumer GPUs such as the RTX 4090~\cite{4090}, restricting its practical use.
%Despite its performance benefits, the GDR-based approach is not used by prominent frameworks like MoonCake~\cite{mooncake} or LMCache~\cite{cachegen} due to two critical limitations. First, it demands immense programming complexity, requiring developers to manage the RDMA network stack and stream synchronization directly from CUDA. Second, and more importantly, its hardware support is limited to expensive datacenter GPUs (e.g., A100/H100) and is unavailable on popular consumer-grade GPUs like the RTX 4090, severely limiting its practical applicability.
%The GPU-driven model suffers from limited hardware support. GDR is a feature exclusive to expensive, datacenter-class GPUs like the NVIDIA A100/H100 and is not available on widely-used consumer GPUs such as the RTX 4090~\cite{}, which severely limits its practical use.

\subsection{The Inefficiencies in RDMA-Based MemPool}
\label{subsec:efficience}

\noindent
While RDMA offers a high-throughput interconnect, it is not a silver bullet for memory disaggregation. Relying on RDMA forces fundamental architectural trade-offs that manifest as critical performance inefficiencies and prohibitive system complexity.

\noindent
\textbf{Performance inefficiencies}. Specifically, the performance inefficiencies stem from three main sources: 
\begin{itemize}
  [itemsep=0pt, parsep=0pt, labelsep=3pt, leftmargin=*, topsep=0pt, partopsep=0pt]

  % \item \textbf{Performance inefficiencies:}
  % \begin{itemize}[itemsep=0pt, parsep=0pt, labelsep=3pt, leftmargin=*, topsep=3pt, label=\textbf{(\arabic*)}]

% \begin{itemize}[itemsep=0pt, parsep=0pt, labelsep=3pt, leftmargin=*, topsep=0pt]
% %\item \textbf{(1) Data path: Extra Data Movement.}
% %\item \textbf{Inefficient Data Path via Host Staging.}

\item 
% \noindent
\textit{Indirect host-staged data path}.
The CPU-driven model subverts the "direct" nature of RDMA by forcing all data through a "bounce buffer" in host memory. This creates a costly data path (GPU → Host DRAM → Remote Memory for writes, and the reverse for reads). This mandatory data staging introduces significant latency for data transfers between GPU and memory pool.

%\item \textbf{(2) Control Path: Control Overhead.}
%\item \textbf{(3) Control Path: Synchronization Overhead.}
\item 
% \noindent
\textit{Complex and inefficient control path}.
A fundamental limitation of RDMA-based offloading is the costly synchronization required between different components. The CPU-driven model, for example, requires CPU-GPU coordination. Similarly, the GPU-driven model needs synchronization between its polling and compute Streaming Multiprocessors (SMs).
In either case, this cross-component coordination imposes a significant latency penalty. This stands in sharp contrast to the seamless execution within a single, unified GPU stream~\cite{bam}.
Our microbenchmark on an H20 GPU reveals the severity of this overhead. 
A 16 KB transfer takes 10.55 µs in total, but the actual data movement accounts for only 2.68 µs.
The remaining ~8 µs (nearly 75\% of the total latency) is attributable to synchronization overhead. This cost stems from launching the kernel or  waiting for its completion. In fact, this synchronization penalty alone is almost 3x the duration of the data transfer itself. 
\revisionb{Futhermore, unlike simple bulk transfers, KVCache reads and writes are complex due to differing data layouts between the GPU and the memory pool. For example, a single KVCache block in Qwen-32B (GQA) requires 128 non-contiguous 20KB data transfers. This introduces extra overheads in the control path. }

\item 
% \noindent
\textit{Inefficient resource utilization}.
RDMA's polling-based control path inherently leads to resource waste. In the CPU-driven model, a host thread consumes entire CPU cores simply to poll for network completions. 
This problem is significantly exacerbated in the GPU-driven model. Here, a dedicated polling kernel occupies valuable SMs. These SMs are the GPU's most critical and limited resource. This occupation creates direct contention with inference tasks and also complicates stream synchronization~\cite{liu2024deepseek, bam}.
\end{itemize}

\noindent
\textbf{System complexity.} Beyond performance issues, RDMA-based approaches also suffer from significant system complexity, including:
%The practical adoption of RDMA-based offloading, especially the GPU-driven model, is severely hindered by two layers of complexity.
\begin{itemize}[itemsep=0pt, parsep=0pt, labelsep=3pt, leftmargin=*, topsep=0pt]
\item 
% \noindent
\textit{Non-trivial development complexity}.
RDMA itself presents prohibitive programming complexity. Instead of simple memory operations, developers are burdened with low-level network management. The GPU-driven model further forces developers to manage the RDMA stack within CUDA and orchestrate synchronization between polling and computation streams.

\item 
% \noindent
\textit{Optimization complexity for tiered memory}.
The overheads of RDMA-based memory access lead to substantial latency penalties for remote KVCache operations. 
This latency may overshadow the computational savings from cache reuse, ultimately negating the entire benefit of offloading the KVCache.
To mitigate this performance penalty, developers must manually implement critical optimizations like request batching~\cite{chen2019scalable, kalia2016design} and ensure data consistency for one-sided RDMA through careful management of QP ordering semantics and software checks~\cite {ziegler2023design, del2022rethinking}.
This further forces systems to adopt complex, cache-aware scheduling policies that tightly couple scheduling decisions to KVCache locality. 
This locality-driven approach creates significant drawbacks, including high scheduling complexity, poor load balancing, and increased maintenance overhead~\cite{Dynamo, mooncake, zuo2025cloudmatrix}.
% Further, systems are forced to adopt complex, cache-aware scheduling policies. This creates a tight coupling between request scheduling and KVCache locality. As a result, requests must be routed to specific nodes that hold the relevant cache data. 
% Systems like NVIDIA Dynamo~\cite{Dynamo} and MoonCake~\cite{mooncake,mooncake} demonstrate the negative consequences. This design introduces significant scheduling complexity, hurts load balancing, and imposes a heavy burden on system design and maintenance~\cite{zuo2025cloudmatrix}.
\end{itemize}
\begin{figure}[t!]
  \centering
    \includegraphics[width=0.9\linewidth]{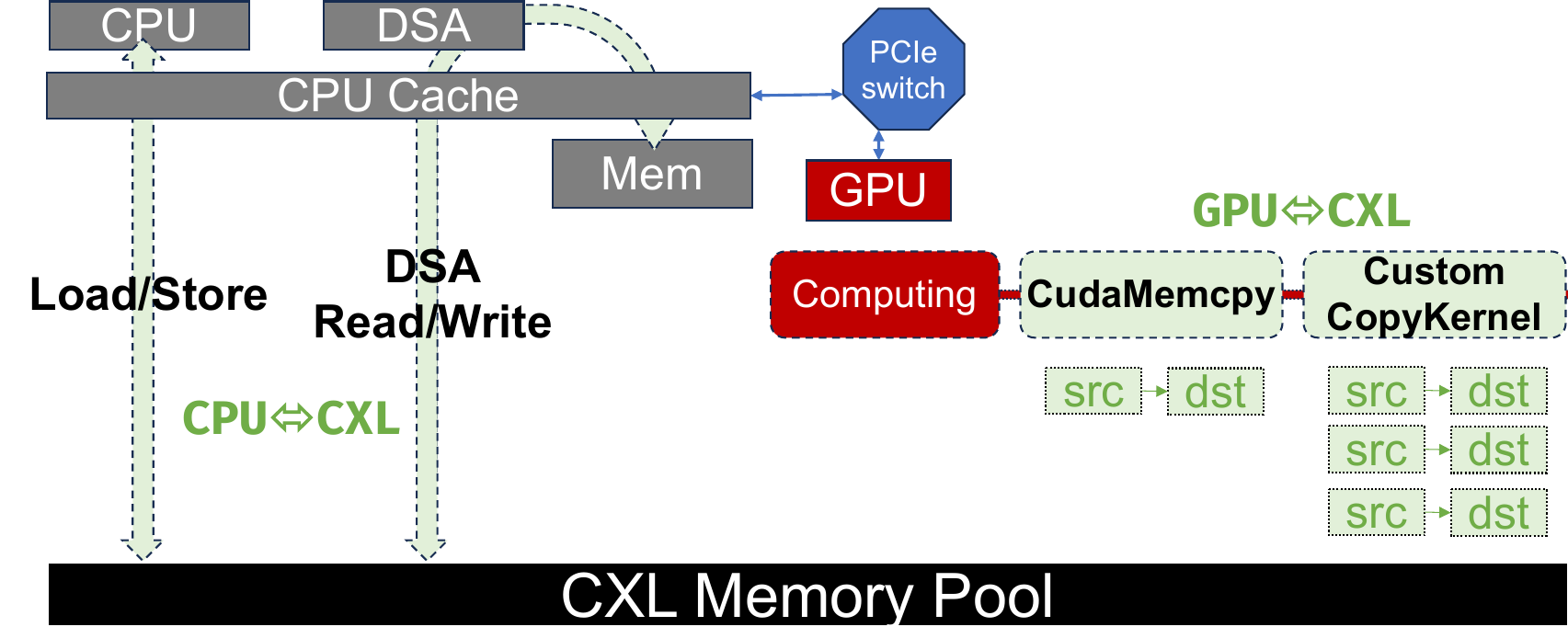}
  \vspace{-0.1cm}
  \mycaption{pic:interface}{Data access interfaces of \sys}{}
  \vspace{-0.2cm}
\end{figure}

\section{\revisionc{\sys: A CXL-Based Memory Architecture}}

%\subsection{\sys: A CXL-Based Memory Architecture}
%\subsection{GPU Clusters with CXL MemPool}
\label{gpucxl}
To overcome RDMA's limitations, we propose \sys, an architecture that leverages the industry's first production-ready CXL switch (XConn XC50256~\cite{xconn}) to build a scalable, shared memory pool. This allows GPUs to access pooled memory with simple load/store operations, solving the performance bottlenecks and programming complexity of the RDMA approach. In this sections, we first give an overview of \sys's architecture and then show its inhenrent advantages.

\subsection{Overview}

As shown in Figure~\ref{fig:cxlM}, this architecture replaces the four dedicated RDMA NICs with two PCIe/CXL adapters, and the hardware deployment of \sys is shown in \autoref{pic:arch}.
Each server has two CPU sockets (NUMA architecture), and each socket connect to the CXL switch via a PCIe 5.0 x16 PCIe/CXL adapter.
The CXL memory pool itself comprises a switch node and a separate memory box. 
At its core, the switch is equipped with two chips (XConn XC50256). Each chip provides 2 TB/s of forwarding capacity over 256 PCIe 5.0 lanes. These lanes are typically split evenly between CXL memory devices and compute servers.
The underlying CXL switch connects up to 16 servers to an 8 TB memory pool with 1 TB/s of total bandwidth. This connectivity enables \sys to support concurrent multi-host access through its internal address mapping and forwarding logic.

\subsection{The Inherent Advantages of \sys}
\label{subsec:beluga}
By altering the memory access paradigm from a network protocol (RDMA) to a memory-semantic interface (CXL), \sys offers several straightforward advantages over RDMA-based approaches (\autoref{pic:interface}):

\noindent
\textbf{Improved performance.}
As shown in \autoref{pic:interface}, \sys provides standard data access interfaces for both CPUs and GPUs. For the CPU, \sys supports (1) direct memory access via \texttt{load/store} instructions and (2) hardware-accelerated transfers using the Intel Data Streaming Accelerator (DSA), a DMA engine available in Intel's Sapphire Rapids CPUs. For the GPU, it supports (1) direct peer-to-peer (P2P) transfers via the \texttt{cudaMemcpy} API and (2) fine-grained, non-contiguous access through custom CUDA kernels. 
These methods provide a straightforward performance benefit compared with RDMA, in both data and control paths.

\begin{itemize}[itemsep=0pt, parsep=0pt, labelsep=3pt, leftmargin=*, topsep=0pt]
\item
\textit{In the data path}, \sys allows the GPU to access the global memory pool directly, eliminating the multi-step data path and bounce buffers required by CPU-driven RDMA, which significantly reduces latency. 

\item
\textit{In the control path}, the data transfer kernels in \sys integrate seamlessly into the GPU's native CUDA stream. This unification of control eliminates the expensive cross-component synchronization inherent in both CPU-driven and GPU-driven RDMA, thereby removing the overheads associated with external CPU coordination or internal GPU polling.
\end{itemize}

\noindent
\textbf{System simplification.}
Beyond performance gains, \sys also introduces fundamental system-level simplifications. These advantages manifest in a more accessible programming model, streamlined memory management, and reduced hardware cost.
% \begin{enumerate}[itemsep=0pt, parsep=0pt, labelsep=3pt, leftmargin=*, topsep=3pt, label=\textbf{(\arabic*)}]
% %\item \textbf{(1) Data path: Extra Data Movement.}
% %\item \textbf{Inefficient Data Path via Host Staging.}
% \item 
% \textit{Simplified Programming Model.}
\begin{itemize}[itemsep=0pt, parsep=0pt, labelsep=3pt, leftmargin=*, topsep=0pt]
\item
\textit{Programming model.}
The data access interfaces are analogous to those used for local DRAM. Consequently, developers are freed from managing the low-level network stack, complex work-request preparations, and performance optimizations required by RDMA-based memory pools. Furthermore, the elimination of cross-component synchronization overhead for GPUs simplifies the coordination between data transfers and computation.
% \item 
% \textit{Simplified Memory Management.} 

\item
\textit{Memory management.} \sys provides a unified address space with simplified space control.
During startup, the BIOS on each host detects the attached CXL devices and reserves a contiguous physical address space for them. \sys then leverages this foundation by having each host manage the CXL memory pool in Direct Access (DAX) mode. In this mode, the CXL memory is exposed as a block device, allowing user-space processes to use the \texttt{mmap()} to map the entire memory region directly into their virtual address spaces.
This direct-mapping mechanism is the key to both resource partitioning and data sharing. Applications built on \sys can either assign different memory offsets to different hosts to achieve logical resource partitioning, or have multiple hosts map the same region to enable efficient data sharing. Crucially, this approach provides a unified memory view across all servers, facilitating simple and efficient memory management.
% \item 
% \textit{Reduced Hardware Cost.} 

\item
\textit{Hardware cost.}
Further, high-speed networking hardware, such as 400Gbps NIC, is often over-provisioned for the typical bandwidth demands of LLM inference~\cite{mooncake}. Replacing expensive RDMA NICs with more cost-effective CXL components can therefore significantly reduce the total cost of ownership. \revisiona{As shown in Table~\ref{tbl:costx}, the CXL-based approach reduces the cost of host interface cards and switches at the device level.
For memory devices, unlike traditional DIMM memory in the host, CXL memory devices can use lower-density chips, reducing the cost per GB. At the system level, CXL memory pools separate memory from CPU resources, allowing multiple servers to share memory and improve utilization in cloud environments.}
\end{itemize}

% Compared with an RDMA-based memory pool, this approach significantly reduces the number of CPUs required in memory nodes and lowers the overall power consumption. 
\begin{table}[h]
 \centering
  \tblcaption{tbl:costx}{\revisiona{Hardware Cost Analysis}}{}

  \resizebox{0.9\linewidth}{!}{
  \begin{threeparttable}[t]
  \centering
    \begin{tabular}{c|cc}
      \toprule      
      & \textbf{RDMA-based} & \textbf{CXL-based} \\ 
      \midrule
      Interface on Host & CX-7 (2$\times$200Gbps) & PCIe/CXL Adapter\\  
      PCIe Lanes  & x16 & x16 \\ 
      Price & \$1,745 & \$210 \\
      \midrule
      Switch & Mellanox RoCE Switch & XConn CXL Switch \\  
      Ports & 40$\times$200Gbps & 32$\times$PCIe5.0 x16 \\ 
      Price & \$16,000 & \$5,800\tnote{*}\\ 
      \$ / (64GB/s) & \$800 & \$218.75 \\
      %\midrule
      %Memory & 64GB DIMM & 256GB CXL Memory\\  
      %Price & \$600 & \$1,800 \\
      %\$ / GB & \$9.38 & \$7.03 \\
      \bottomrule
    \end{tabular}

\begin{tablenotes}
     
  \item[*]B1 sample price, and final price subject to change.
\end{tablenotes}
\end{threeparttable}
}
\vspace{-0.2cm}
\end{table}

\revisionb{In summary, for a single-rack memory pool, CXL demonstrates clear advantages over RDMA by offering an optimal balance among performance, cost-effectiveness, and resource utilization.}

\section{Characterization and Optimization of \sys}
\label{sec:featofsys}
%\section{Software Design and Optimization of \sys}

\label{sec:characterizing}

\noindent
While CXL provides a simple memory-like interface, the poorly understood performance of its CXL 2.0 hardware, particularly for direct GPU-to-memory access, hinders the development of software that can unlock its full potential.
%While CXL provides a simple memory-like interface, the foundational software must carefully consider the hardware's unique characteristics to unlock its full performance potential. However, the performance of emerging CXL 2.0 hardware, particularly for direct GPU-to-memory data paths, remains insufficiently characterized. 
To address this gap, this section presents a comprehensive performance analysis of all data paths in \sys, yielding a set of practical optimizations summarized in \autoref{tbl:all}. The analysis proceeds in three parts:
First, we design and evaluate three software-based coherence methods, providing clear optimization guidelines for developers (\S\ref{subsec:non-cc}).
Second, we analyze different I/O transfer methods and identify the optimal approach for latency-sensitive workloads  (\S\ref{subsec:basic-perfor}).
Third, we evaluate and identify the bandwidth bottleneck in \sys and propose two effective solutions to scale performance for bandwidth-intensive workloads (\S\ref{subsec:throughput}).
All experiments are validated on a platform using commercially available hardware, with detailed specifications in \autoref{tbl:env}.

\begin{table}[h!]
\vspace{-0.1cm}
  \tblcaption{tbl:env}{Experimental setup}{}
  \resizebox{0.90\linewidth}{!}{
    \begin{tabular}{c|c}
      \toprule      
      OS& Ubuntu 22.04, kernel 6.2.0-1015\\ 
      CPU& 2 $\times$ Intel(R) Xeon(R) Platinum 8575C  \\  
      L3 Cache & 640 MiB (320 MiB per CPU) \\ 
      DRAM & 2~TB (32 $\times$ DDR5 4800~MT/s 64~GB)\\
      PCIe & 4 $\times$ PCIe Switch, 5.0 \\
      GPU & 8 $\times$ H20 \revisionx{(96~GB)} \\
      \midrule
      NICs / Server & 4 $\times$ ConnectX-7 (Dual-port 200Gbps) \\
      RDMA MemPool & 4~TB (2 $\times$ GPU Servers) \\ 
      \midrule
      PCIe/CXL Adapter / Server & 2 $\times$ PCIe 5.0 x16 \\
      CXL MemPool & 8~TB (32 $\times$ DDR5, 4800~MT/s, 256~GB)\\
      \bottomrule
    \end{tabular} 
  }
  \vspace{-0.1cm}
\end{table}

\begin{table*}[]
  \tblcaption{tbl:all}{Performance optimizations in \sys}{}
  \resizebox{1\linewidth}{!}{
    \begin{tabular}{c|l|c}
      \toprule      
      \textbf{Aspect} & \multicolumn{1}{c}{\textbf{Optimizations}} & \textbf{Applicable To} \\
      \midrule
      \multirow{3}{*}{Cache Coherency (\S\ref{subsec:non-cc})} 
      % & \textbf{L1.} Shared CXL devices lack hardware cache coherency  support. & - \\
      & \textbf{O1.} For CPU access, use \texttt{ntstore} for writes and invalidate CPU cache before reads. & CPU \texttt{load/store} \\
      & \textbf{O2.} For CPU DSA, set memory as Uncachable. & CPU DSA \\
      & \textbf{O3.} For GPU access, set memory as Uncachable and disable DDIO. & GPU\\
      \midrule
      \multirow{3}{*}{Latency (\S\ref{subsec:basic-perfor})} 
      % & \textbf{A1.} For small I/O sizes, CXL offers significantly lower latency than RDMA (2.3$\times$--3.3$\times$). & - \\
       % & \textbf{A2.} For large I/O sizes, CXL maintains a 10\%--20\% latency advantage over RDMA. & - \\
      & \textbf{O4.} Use direct \texttt{load/store} for small I/Os ($<$ 4~KB) and use DSA for larger transfers. & CPU\\
      & \textbf{O5.} Launch kernels asynchronously using CUDA streams to hide the launch latency. & GPU\\ 
      & \textbf{O6.} For GPU transfers ($<$ 24~KB) on Uncachable memory, use customized copy kernel. & GPU \\
      \midrule
      \multirow{3}{*}{Bandwidth (\S\ref{subsec:throughput})} 
      % & \textbf{L2.} Peak bandwidth of a single CXL Interface Card is limited by Root Complex (RC). & - \\
      & \textbf{O7.} Support direct GPU-to-CXL switch connections in future architecture. & GPU \\
      & \textbf{O8.} Use more PCIe/CXL adapters for scalable bandwidth. & - \\
      & \textbf{O9.} Interleave data across multiple CXL memory devices. & - \\
      \bottomrule
    \end{tabular}
  }
\end{table*}

\subsection{Data Sharing over Non-Coherent CXL}
% \subsection{Data Sharing via CXL w/o Hardware Cache Coherence}
\label{subsec:non-cc}
%A foundational challenge in designing systems with CXL 2.0 memory pooling stems from its memory consistency model. The CXL 2.0 specification, while enabling memory disaggregation via switches, does not support hardware-managed cache coherency across different host nodes. A CXL switch fabricates a shared memory space, but each host's CPU interacts with it through its own private cache hierarchy. As depicted in a typical multi-host setup (see Figure X), a write operation from HostA is cached in its local Last-Level Cache (LLC). This write is not automatically propagated to HostB's cache, nor is HostB's cached copy of the same address automatically invalidated. 

%The absence of hardware-managed cache coherency is a foundational challenge in CXL 2.0 memory pooling. While a CXL switch exposes a unified memory address space, each host maintains a private, non-coherent cache hierarchy. 
%Consequently, a write by one host remains local to its cache hierarchy (L1/L2/L3)  and is not propagated to peer hosts, creating the risk of observing stale data. 
%For KVCache, this could mean an LLM generating output based on an incorrect context. 
%This lack of hardware coherency shifts the burden of data consistency to the software,  which must then explicitly perform cross-host synchronization.

\noindent
\textbf{Goals and methodology.}
While CXL.cache and CXL.mem in CXL 2.0 support coherent memory sharing between host CPUs and devices like accelerators or memory expanders, they do not support cache coherence across multiple host CPUs. While CXL switches present a logically unified memory address space to host processors, each computing node maintains an independent cache hierarchy (L1/L2/L3) that operates in isolation from the other nodes. 
As a result, a write by one host remains confined to its local cache and is not automatically propagated to remote hosts. 
In the KVCache, this deficiency leads to inaccurate, inconsistent, or incoherent output. This occurs when one GPU reads a stale cache entry that is being concurrently updated by another GPU.
% The CXL.cache protocol supports coherence between a host and a device, but CXL 2.0 does not extend coherence across multiple hosts connected via a CXL switch. 
%As a result, systems built on CXL 2.0 require explicit software or protocol-level mechanisms to maintain consistency across host caches when accessing shared memory.
Without hardware-supported multi-host coherence, the burden of maintaining data integrity shifts to software. This necessitates explicit cross-node synchronization protocols to coordinate all cache state transitions.

% \vspace{0.1cm}
\revisiona{Our system focuses on optimizing KVCache sharing mechanisms, enabling a single writer to insert a KVCache block while allowing multiple readers to access it, thereby eliminating redundant computations. 
Targeting this scenario,} we propose three software-managed methods to enforce consistency for the writer-side and two methods for the reader-side. We then conduct a series of experiments to measure the latency and to select the optimal method for different access operations. 
% Based on the experimental results, we select the lowest-latency 
% method in the subsequent experiments.

\noindent
\textbf{Writer: ensuring data reaches CXL memory.}
To ensure a host's writes are visible to other hosts, the data must be flushed from its private cache hierarchy to the CXL memory. There are three methods to achieve this:
% \begin{itemize}
% [itemsep=0pt, parsep=0pt, labelsep=3pt, leftmargin=*, topsep=0pt, partopsep=0pt]

\noindent
%\underline{Using uncacheable (UC) memory and disabling DDIO.} 
%The most direct approach is to configure the physical memory regions corresponding to the CXL pool as Uncacheable (UC). 
%In our enviroment, we set the Memory Type Range Registers (MTRRs) to Uncachable~\cite{mtrr}. Every load or store from CPU to such a region bypasses the CPU cache hierarchy and directly translates into a PCIe/CXL transaction. %Note that this method does not work for GPU. 
%For Device-to-Host (D2H) memcpy from GPU, as also as other write operations for devices, the direct approach is to disable Data Direct I/O (DDIO). By default, DDIO directs inbound traffic from devices like GPUs and NICs into the CPU's Last-Level Cache (LLC) for high performance. Disabling DDIO alters this behavior, forcing D2H memcpy traffic to bypass the LLC and write directly into the CXL memory pool. 
%\underline{Coarse-Grained Cache Bypassing.}

\begin{itemize}
  [itemsep=0pt, parsep=0pt, labelsep=3pt, leftmargin=*, topsep=0pt, partopsep=0pt]
\item 
\textit{Static uncacheable configuration.}
A straightforward approach for enforcing memory coherence in CXL-based systems is to configure the CXL memory regions as uncacheable, by setting the corresponding memory type in the Memory Type Range Registers (MTRRs)~\cite{mtrr}. Consequently, any CPU-initiated write, whether via a standard \texttt{store} instruction or a DSA transfer, bypasses the host's cache hierarchy and is sent directly to the \sys CXL memory pool.
Symmetrically, for Device-to-Host (D2H) memcpy from GPU, we disable Data Direct I/O (DDIO)~\cite{ddio}. By default, DDIO directs inbound traffic from devices like GPUs and NICs into the CPU's Last-Level Cache (LLC) for high performance. Disabling DDIO alters this behavior, forcing D2H memcpy traffic to bypass the LLC and write directly into the CXL memory pool. 

\item 
%\underline{Flushing cache after write.} 
\textit{Fine-grained cache flushing after write.} 
This approach caches data but relies on software to explicitly flush modified cache lines to memory using CPU instructions like \texttt{CLFLUSH}, \texttt{CLFLUSHOPT} and \texttt{CLWB}~\cite{clflush}.
% \texttt{CLWB} is often preferred as it avoids invalidating the cache line to benefit local re-reads. 
However, these methods impose high instruction overhead, which scales linearly with the data size, as it uses per-cache-line operations. 
%The cost scales linearly with the data volume, as flushing a large KVCache requires executing an instruction for each of its thousands of cache lines.

\item 
\textit{Fine-grained bypassing-cache write without cache flushing.} 
To eliminate the overhead of explicit flushing, we can leverage some bypassing-cache instructions in the CPU. CPU-originated writes can use non-temporal store instructions (\texttt{ntstore}),%~\cite{}, 
which bypasses the cache hierarchy and writes data into the CXL memory pool directly. Similarly, Intel DSA provides a cache bypass flag, enabling DMA operations to write their payloads directly to memory without populating the CPU Cache.  
\end{itemize}
% However, a critical prerequisite underpins the effectiveness of these cache-bypassing techniques: \textbf{the target memory region must not be present in the cache prior to the non-temporal store operation}.   
% \end{itemize}

\noindent
\textbf{Reader: ensuring fresh data from CXL memory.}
Similarly, to prevent reading stale data in local cache, the system should either configure the memory region as uncacheable or explicitly invalidate the target cache lines before the read operation.

% \begin{itemize}
% [itemsep=0pt, parsep=0pt, labelsep=3pt, leftmargin=*, topsep=0pt, partopsep=0pt]
\begin{itemize}
  [itemsep=0pt, parsep=0pt, labelsep=3pt, leftmargin=*, topsep=0pt, partopsep=0pt]
\item 
\textit{Static uncacheable configuration. } 
For CPU reads, the uncacheable attribute ensures correctness by forcing every load to access the remote CXL memory pool. 
Correct GPU access to CXL memory requires the uncacheable attribute, even for device-initiated transfers like \texttt{cudaMemcpy}. Although the GPU operation itself does not populate the CPU cache, concurrent or prior CPU activity can create cached copies of the data. Without this attribute, the GPU might read stale data from the CPU cache, leading to data inconsistency. 
\item 
\textit{Fine-grained flushing cache before read.} 
%This method is to proactively invalidate any potentially stale data from the local cache before issuing the read. Instructions like CLFLUSH can serve this purpose (as they evict and invalidate), while CLWB can not. Software must meticulously track which memory regions might be stale and issue targeted CLFLUSH instructions on those cache lines before reading. This method imposes significant software complexity and instruction overhead for consistency. 
This approach ensures read correctness by proactively invalidating stale cache lines with \texttt{CLFLUSH} instructions before the read operation. Note that the \texttt{CLWB} does not work for this purpose. These flush instructions suffer from high overhead as they operate on a per-cache-line granularity.

\end{itemize}

\revisiona{The above designs in writer or reader is simple: setting up cache states and disabling DDIO are one-time steps, and cache flushing is as simple as read/write. 
In contrast, achieving cache coherence in RDMA requires more complex management of asynchronous completions and queue pair (QP) ordering.}

% \noindent
% \textbf{The most efficient method.}

%\vspace{1em}

\noindent
\textbf{(\textbf{Exp \#1}) Performance characterization.} We evaluate the latency of 16~KB read/write and the results are shown in \autoref{tbl:cc}. 
For write operations, the results reveal a clear performance hierarchy. 
For CPU-initiated store, non-temporal stores (\texttt{ntstore}) are the most efficient method, achieving a latency of 2.41~$\mu s$. They bypass the cache and also avoids the explicit flush overhead. In contrast, standard writes followed by CLFLUSH-family instructions are slower (8.50~$\mu s$). Using uncacheable memory is prohibitively slow (281.56~$\mu s$), as each \texttt{store} instruction stalls the CPU pipeline while waiting for the entire CXL access to complete. 
For the DSA, both using uncacheable memory and the cache-bypassing flag yield nearly identical, top-tier performance, as the DSA is not hindered by CPU pipeline stalls associated with uncacheable memory access. 
For GPU D2H transfers, disabling DDIO to bypass the cache is more effective (9.14$\mu s$) than relying on a subsequent, slow CPU flush (11.06$\mu s$).
For read operation, we observe that CPU \texttt{load}s from uncacheable memory are prohibitively slow 
(166.49~$\mu s$). Consequently, the only viable method for CPU \texttt{load} is to flush the cache before the read, which achieves a much lower latency of 5.98~$\mu s$. 
Similar to its write performance, the CPU DSA read and GPU H2D copy perform better when reading directly from uncacheable memory.

\begin{table}[t]
  \centering
  \tblcaption{tbl:cc}{Latency of cache coherency methods ($\mu s$)}{\mdseries \revisionc{Bold entries highlight the optimal mechanism for each operations. (\textbf{Exp \#1})}}
\begin{threeparttable}[t]
%\begin{table}[]
  % \caption{Lowest-latency cache coherency methods (~$\mu s$)}
  % \label{tbl:cc}
\resizebox{0.9\linewidth}{!}{
  % \begin{threeparttable}

\begin{tabular}{c|c|c|c}
      \toprule      
      \textbf{Write Direction} & \multicolumn{2}{c|}{\bf CPU $\Rightarrow$ CXL }  & \textbf{GPU $\Rightarrow$ CXL }\\
      Operation (16~KB) & Store & DSA Write & Custom Kernel~\tnote{*}\\
      \midrule
      UC Mem / Disable DDIO & 281.56  & \textbf{1.69}  & \textbf{9.14} \\  
      CLFLUSH after Write & 8.50  & 3.64 &  11.06 \\
      Bypassing-Cache Write & \textbf{2.41}  & \textbf{1.76}  & - \\
      % disable ddio & - & -  & \textbf{9.14us} \\
      \midrule
      \midrule
      \textbf{Read Direction} & \multicolumn{2}{c|}{\bf CPU $\Leftarrow$ CXL }  &\textbf{GPU $\Leftarrow$ CXL}\\
      Operation (16~KB) & Load & DSA Read & Custom Kernel~\tnote{*}\\
      \midrule 
      UC Mem & 166.49 & \textbf{2.12} & \textbf{10.55} \\
      CLFLUSH before Read & \textbf{5.98}  & 4.84 & 16.81  \\
      % Avoiding CPU Read or Uncachable memory & - & -  &  \\
      \bottomrule
\end{tabular} 
}
  \begin{tablenotes}
      \item[*] The result of custom kernel includes the time of kernel launch.
  \end{tablenotes}
  % \end{threeparttable}
\end{threeparttable}
\end{table}
% 

% Configuring the CXL source memory as Uncacheable (UC) is the most efficient method for GPU Host-to-Device transfers (10.55\,\textmu s), outperforming CPU-managed cache flushing (16.81\,\textmu s) by avoiding costly intervention. 

% \begin{mybox}
\noindent
\textbf{Optimizations.} Current CXL 2.0 switches do not support host-to-host hardware cache coherency, a limitation that requires software-managed cache coherency. Optimal performance for data sharing is achieved by applying specific cache management strategies based on the operation's initiator. There are three design hints: 
for CPU store/load, non-temporal stores should be used for writes, and a CLFLUSH should precede loads \textbf{(O1)}. For the Intel DSA engine, operating on uncacheable memory is the best practice for both reads and writes \textbf{(O2)}. Finally, for GPU memcpy operations, disabling DDIO while utilizing uncacheable memory offers the highest efficiency \textbf{(O3)}. \revisiona{Looking ahead, transparent cross-host cache coherence via CXL 3.0 switches could remove the need for software-based synchronization. But CXL 3.0 coherence is still evolving; its region size and guarantees are unclear and may be limited due to cost and complexity. 
Our software-based designs are therefore necessary for non-coherent regions. When hardware coherence is available for small regions, we can use it to further simplify and accelerate our software protocol.}
% \end{mybox}

% \begin{figure}[]
%   \begin{center}
%     \includegraphics[width=\linewidth]{pic/perf1.png}
%   \end{center}
%   \mycaption{pic:perf1}{4MB I/O Latency}{} 
% \end{figure}

% The source and destination addresses are randomized within a 64\,GB space for local memory and a 2\,TB space for the CXL pool to avoid caching effects.
% Each operation was tested over one million iterations to obtain a stable average latency. 

% \begin{figure} 
%   \centering
%   \subfloat[Write Latency]{\includegraphics[width=0.5\linewidth]{fig/w-small.pdf}\label{fig:w-c}}
%   \hfill
%   \subfloat[Read Latency]{\includegraphics[width=0.5\linewidth]{fig/r-small.pdf}\label{fig:r-c}}
%   \mycaption{pic:cpu}{(Exp \#2) CPU $\Leftrightarrow$ Remote Memory Pool}{}
% \end{figure}

% For instance, at 1 KB, a pure CPU-based Load/Store takes only 0.98 µs, whereas a DSA-initiated transfer to CXL memory requires 1.19 µs. This is attributable to the overhead of programming the DMA engine, which outweighs the benefits of offloading for small data chunks.
% This advantage grows with size; at 64\,KB, the DSA's latency (3.38\,\textmu s) is nearly 3x lower than the CPU copy (9.4\,\textmu s), highlighting the indispensability of DMA for efficient, moderately-sized data movement.

\subsection{Latency Optimization}
\label{subsec:basic-perfor}

\noindent
\textbf{Goals and methodology.} 
The performance of CXL memory access for both CPUs and GPUs depends heavily on the specific operation and access size.
We therefore begin by characterizing the latency of four fundamental operations in \sys to identify their optimal use cases. 
Furthermore, we compare \sys against local memory and RDMA-based memory pools as baselines to highlight the scenarios where it offers the most significant performance benefits.
%Furthermore, to provide a broader performance context, we include results from local memory and an RDMA-based memory pool as baselines. This comparison clarifies the scenarios where \sys offers the most significant performance benefits.

\begin{figure}[t!]
  \centering
  \vspace{-0.2cm}
  \subfloat[CPU $\Rightarrow$ Memory Pool]{\includegraphics[width=0.5\linewidth]{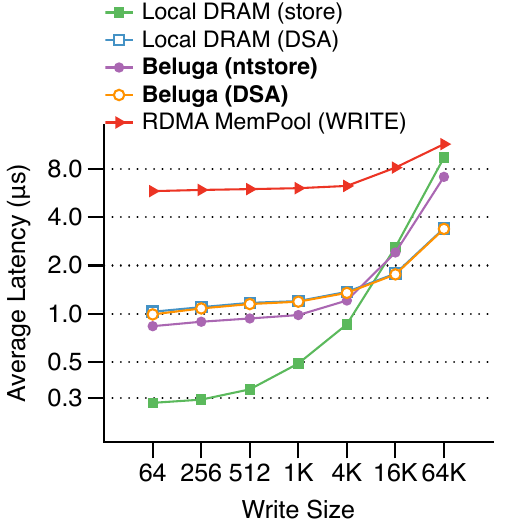}\label{fig:w-c}}
  \hfill
  \subfloat[CPU $\Leftarrow$ Memory Pool]{\includegraphics[width=0.5\linewidth]{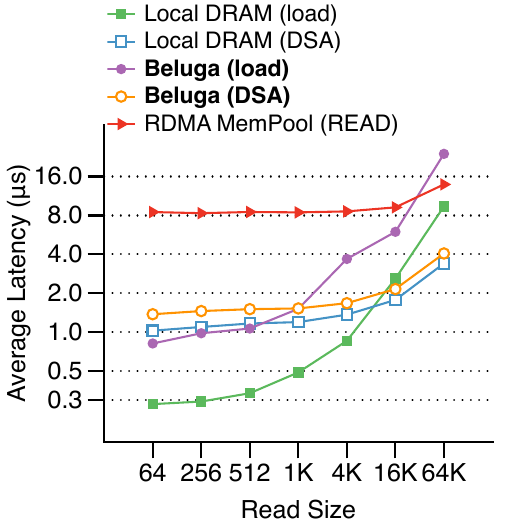}\label{fig:r-c}}
  \hfill
  \vspace{-0.2cm}
  \subfloat[GPU $\Rightarrow$ Memory Pool]{\includegraphics[width=0.5\linewidth]{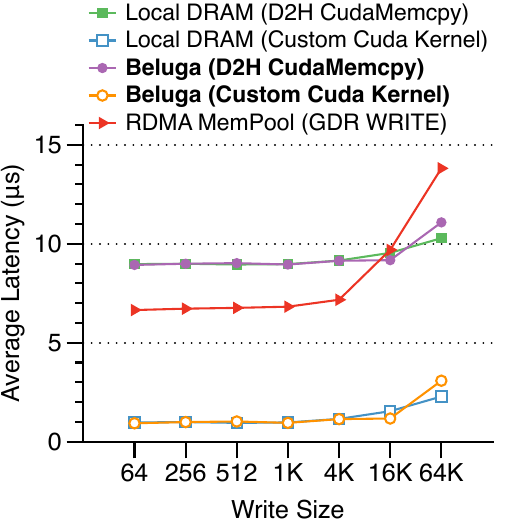}\label{fig:w-g}}
  \hfill
  \subfloat[GPU $\Leftarrow$ Memory Pool]{\includegraphics[width=0.5\linewidth]{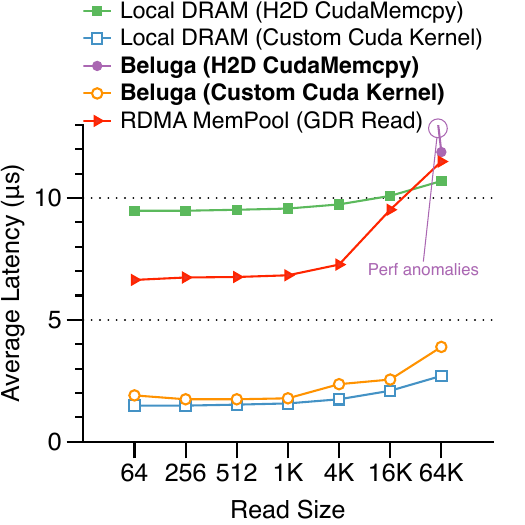}\label{fig:r-g}}
  \vspace{-0.1cm}
  \mycaption{pic:gpu}{Latency between CPU/GPU $\Leftrightarrow$ remote memory pool (\textbf{Exp \#2})}{\mdseries \revisionc{Local memory and RDMA memory pools are included as baselines. This experiment not only demonstrates the performance differences among various CXL access methods but also reveals that CXL memory pooling achieves latency characteristics comparable to local memory pools, significantly outperforming RDMA-based memory pooling in terms of latency.}}

\end{figure}

\noindent
\textbf{(\textbf{Exp \#2}) Performance characterization.}
\autoref{pic:gpu} presents the I/O operations between CPU/GPU and memory pool. Unless otherwise specified, all latency benchmarks use a queue depth of one (QD=1) to measure single-request latency. Our key observations are as follows: 
 
% \noindent
% \textbf{(Exp \#3) Latency of Small I/O.} 

First, for GPU access, direct CXL-to-GPU data transfers are highly competitive with traditional CPU-to-GPU paths (Figure.\ref{fig:w-g} and \ref{fig:r-g}), establishing the CXL pool as a viable data source for the GPU. We find that at a 64\,KB transfer size, the latency of a CXL-to-GPU copy is 11.73\,µs, which is remarkably close to the 10.32\,µs latency of a conventional CPU-to-GPU copy. This result indicates that CXL memory can serve as a primary data source for GPU workloads with negligible latency overhead.

Second, a clear performance trade-off exists between CPU-based copies and DMA-based transfers for moving data to CXL (Figure.\ref{fig:w-c} and \ref{fig:r-c}). For I/O sizes up to 4 KB, direct CPU load/store operations exhibit lower latency than offloading the transfer to the DMA engine. 
The benefit of DMA parallelism begins to outweigh its setup overhead for transfers larger than 4\,KB. The performance crossover occurs at 16\,KB, where the Intel DMA engine DSA outperforms the CPU-based copy. 

% Second, for CPU access to memory pool, CXL demonstrates a clear latency
% advantage over RDMA. For 64\,KB transfers, CXL writes are 3.3$\times$ faster than RDMA writes, and similarly CXL reads exhibit a 2.3$\times$ performance advantage. 

Third, our analysis shows that the poor latency in CUDA Memcpy Kernel is dominated by software overhead, rather than the physical data movement time (Figure.\ref{fig:w-g} and \ref{fig:r-g}). This overhead stems primarily from the kernel launch latency within the GPU driver stack. For instance, while the CUDA Memcpy Kernel for a 16 KB H2D (Host to Device or CPU memory to GPU memory) transfer is 10.55 µs, profiling with NVIDIA Nsight Systems (nsys) reveals that the actual data transfer on the GPU completes in only 2.68$\mu s$. 
\revisionb{Therefore, we recommend using custom data copy kernels to combine multiple copy tasks in a single kernel. This method avoids kernel launch and CPU synchronization overhead. As a result, it achieves better latency compared to both CUDA Memcpy Kernel and GDR.}

% The "poor performance" of Beluga referred to the CPU-centric data path (CUDA Memcpy Kernel), which involves overheads of kernel launches. 

Finally, as shown in Figure.\ref{fig:r-g}, we observe that the standard cudaMemcpy for H2D transfers: when the source memory is Uncachable, performance degrades dramatically for transfers smaller than 24 KB, taking approximately 1.23 ms (values omitted from the Figure.\ref{fig:r-g} for clarity of scale). We hypothesize that the CUDA runtime leverages CPU-based instructions to optimize small transfers, a strategy that is ill-suited for Uncachable memory. Consequently, to achieve efficient H2D data movement for transfers under 24 KB from Uncachable CXL memory, it is necessary to use a custom CUDA memcpy kernel. 

% \noindent
% % \textbf{(Exp \#4) Latency of Large I/O.}
% We evaluate the performance of large I/O operations (larger than 1 MB), with results shown in \autoref{tbl:large}. For CPU access, we use DSA due to its better performance in large sizes.
% Our analysis yields two key findings.

% First, from the CPU's perspective, the CXL Memory Pool offers write performance highly competitive with local memory for moderately sized transfers. The 4\,MB write latency to the CXL pool (133.60\,µs) is nearly identical to that of writing to local memory via Intel's DSA (134.35\,µs). While an overhead emerges for larger transfers, CXL still shows a notable advantage over RDMA for smaller-scale large I/O, such as reducing write latency by 34\% at the 1\,MB size. For very large 1\,GB transfers, the performance of CXL and RDMA pools converges, with RDMA even showing a slight edge in write latency.

% Second, for the GPU, the CXL Memory Pool consistently serves as a more efficient secondary memory tier than the RDMA pool. Across all tested sizes, CXL delivers lower latency for both reads and writes. For instance, at the 1\,GB transfer size, CXL's read and write latencies are approximately 11\% and 10\% lower than RDMA's, respectively. This sustained performance advantage underscores CXL's suitability for enabling GPUs to efficiently perform out-of-core operations on datasets exceeding their local memory capacity.
% \input{sec/large-table.tex}
% \begin{mybox}

\noindent
\textbf{Optimizations.} 
%\sys provides low latency that is competitive with local DRAM.
\sys delivers latency that is competitive with local DRAM. This key characteristic enables the creation of a flat, unified memory pool, with the potential to eliminate the need for complex, multi-tiered memory hierarchies.
Furthermore, we propose the following three optimizations to leverage this low latency. First, the CPU should use direct load/store for small I/Os (< 4 KB) and DSA for larger transfers \textbf{(O4)}. Second, when moving data between the GPU and CXL memory pool, tasks should be combined into a single kernel to hide the kernel launch latency \textbf{(O5)}. Third, for data transfers up to 24 KB from CXL memory to the GPU, a custom-implemented CUDA memcpy kernel is recommended \textbf{(O6)}.

% The CXL memory pool consistently delivers lower latency than an RDMA-based pool for both small and large data transfers.
% For small I/O (< 64 KB), it provides a 2.3$\times$-3.3$\times$ latency advantage over RDMA \textbf{(A1)}. 
% Within this regime, direct CPU load/store is optimal for the smallest transfers (< 4 KB), while DSA is more efficient for larger blocks \textbf{(H5)}. For the GPU, direct CXL-to-GPU transfers are nearly as fast as local CPU-to-GPU operations, positioning the pool as an effective memory extension, but the application should launch kernels asynchronously using CUDA streams to hide the launch latency \textbf{(H6)}.
% For large data transfers, the CXL pool reduces latency by up to 34\% for CPU operations and 10-20\% for GPU transfers compared to RDMA \textbf{(A2)}.

% \end{mybox}

% \subsection{Latency}

% \subsubsection{Breakdown of data store and cache invalidation}

% \noindent
% \textbf{Peak IOPS.}

% For instance, data transfers from host DRAM to a GPU are limited by the GPU-PCIe interface itself, achieve about 55.4~GB/s, while data transfers from remote DRAM (RDMA MemPool) to a GPU are limited by the NIC bandwidth, which is only 44.1~GB/s.
% Our baseline measurements with an RDMA NIC show a practical PCIe 5.0 x16 throughput of approximately 44.1 GB/s, which is about 70\% of the theoretical maximum.
\subsection{Bandwidth Optimization}

\label{subsec:throughput}
\noindent
\textbf{Goals and methodology.} We observe two performance anomalies in \sys when we use a single PCIe/CXL Adapter (16 lanes). 
First, it exhibits asymmetric read/write performance. The read bandwidth from CPU to CXL memory pool reaches the expected 46.2 GB/s, which is comparable to that of RDMA. However, the write throughput is limited to 33 GB/s. Second, we observe a significant performance degradation during GPU access to the CXL memory pool, with throughput dropping to 26 GB/s. 
This is substantially lower than the bandwidth of CXL memory controllers and the GPU's own PCIe bandwidth (55.4 GB/s).
The data path from a CPU/GPU to the CXL memory pool is complex, traversing components such as the PCIe Switch, Root Complex, and CXL Switch. Any of these components can become a performance bottleneck, limiting the end-to-end bandwidth. Our approach is to first perform a bottleneck analysis of this path and then present targeted optimizations.

% \begin{table}[htbp]
% \begin{threeparttable}
%   \resizebox{1\linewidth}{!}{
%     \begin{tabular}{c|cccc}
%       \toprule      
%       Device & Root Complex  & CXL Interface Card & CXL Device\\ 
%       \# of lanes &  & x16 & x8 \\ 
%       \midrule
%       Bandwidth & 55.4~GB/s & 49.3~GB/s & 46.2~GB/s\tnote{*} & 22.5~GB/s \\
%       \bottomrule
%     \end{tabular} 
%   }
%   \begin{tablenotes}
%       \item[*] 46.2~GB/s is read bandwidth, while write bandwidth is 33.0~GB/s.
%     \end{tablenotes}
%   \tblcaption{tbl:Throughput}{(Exp \#5) Peak throughput of a single device}{}
%  \end{threeparttable}
% \end{table}

% \begin{table}[htbp]
% \begin{threeparttable}
%   \resizebox{1\linewidth}{!}{
%     \begin{tabular}{c|cccc}
%       \toprule      
%       Device & PCIe Switch & Root Complex P2P & CXL Interface Card & CXL Device\\ 
%       \# of lanes & x16 & x16 & x16 & x8 \\ 
%       \midrule
%       Bandwidth & 55.4~GB/s & 33.0~GB/s, 26.0~GB/s\tnote{*} & 46.2~GB/s & 22.5~GB/s \\
%       \bottomrule
%     \end{tabular} 
%   }
%   \begin{tablenotes}
%       \item[*] 33.0~GB/s is P2P write, 26.0~GB/s is P2P read.
%   \end{tablenotes}
%   \tblcaption{tbl:Throughput}{Peak throughput of a single device}{}
%  \end{threeparttable}
% \end{table}

\begin{figure}[htbp]
  \centering
    \includegraphics[width=0.8\linewidth]{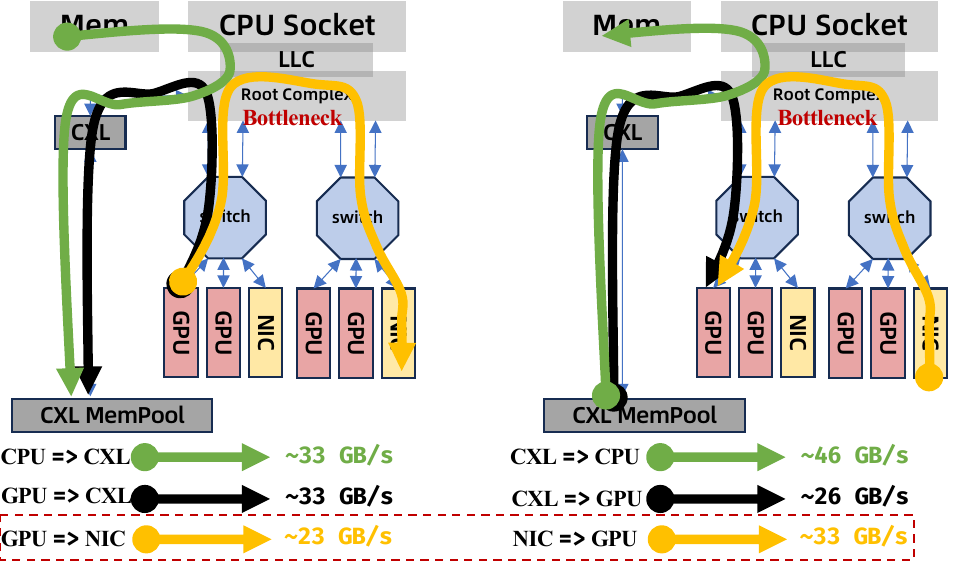}
  \mycaption{pic:bw}{Bandwidth for different CXL access paths}{} 
\end{figure}

\noindent
\textbf{Performance characterization.}
Through the systematic empirical analysis, we identify the CPU's Root Complex (RC) as the primary performance bottleneck in CXL memory pool architectures. 
To verify this conclusion, we construct a dedicated micro-benchmark  (\autoref{pic:bw}). It measures the bandwidth between a GPU and a NIC that are attached to different PCIe switches, forcing the data path to go through the CPU's Root Complex.
The results of this experiment precisely match the performance with CXL memory pool: the PCIe P2P write (i.e., NIC$\Rightarrow$GPU) bandwidth was 33 GB/s, and the P2P read (i.e., GPU$\Rightarrow$NIC) bandwidth was limited to 23 GB/s. 
This indicates that the bottleneck may not be in the CXL memory devices or the CXL links themselves. Instead, performance is limited by the underlying peer-to-peer capabilities of the Root Complex.
To further understand this bottleneck, we conduct additional experiments with multiple GPU-NIC pairs. 
The aggregate bandwidth scales linearly with the number of pairs: two pairs achieve 46 GB/s, approximately double the throughput of a single pair.
It suggests that the limitation lies in per-lane or per-link resources within the Root Complex rather than an overall throughput cap.

Furthermore, we observe that the memory devices themselves can be another bottleneck. Each device supports a bandwidth of 22.5 GB/s. If all workloads from a server are directed to a single device, the bandwidth will be limited by that device.
Therefore, software running on \sys should distribute data across different memory devices to avoid the single-device bottleneck. 
\revisionc{Our current implementation relies on software-based memory interleaving at a 2MB granularity.
The upcoming Intel Granite Rapids (6th-generation Xeon) processors provide hardware support for CXL device interleaving, offering flexible configurations from 256B chunks up to eight-way interleaving.}

\noindent
\textbf{Optimizations.} Based on our observations, we further propose three optimization strategies for bandwidth-sensitive scenarios.
First, future architecture should support direct GPU-to-CXL switch connections. As such, \sys is able to bypass the RC, avoiding the major bottleneck between the GPU and CXL memory \textbf{(O7)}.
Second, bandwidth of a single PCIe/CXL adapter is limited to PCIe5.0 x16. And, for higher bandwidth, the number of PCIe/CXL adapters should be scale with application requirements \textbf{(O8)}.
Third, \sys interleaves data across multiple CXL memory devices. This strategy parallelizes access and helps maximize the aggregated throughput \textbf{(O9)}.
\begin{figure}[t!]
  \centering
  \subfloat[Bandwidth (64B)]{\includegraphics[width=0.333\linewidth]{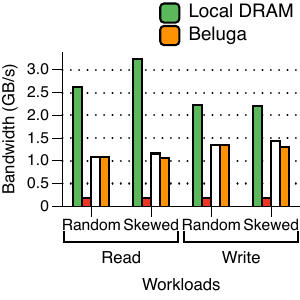}\label{fig:64bw}}
  \hfill
  \subfloat[Median Latency (64B)]{\includegraphics[width=0.333\linewidth]{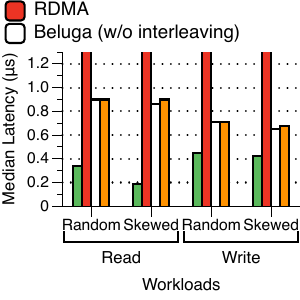}\label{fig:64p50}}
   \hfill
  \subfloat[P99 Latency (64B)]{\includegraphics[width=0.333\linewidth]{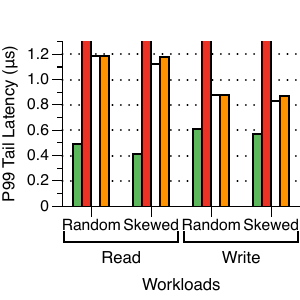}\label{fig:64p99}}
%   \mycaption{fig:64}{Performance under concurrent 64B accesses}{}
% \end{figure}
    \hfill

% \begin{figure}[htbp]
%   \centering
  \subfloat[Bandwidth (16K)]{\includegraphics[width=0.333\linewidth]{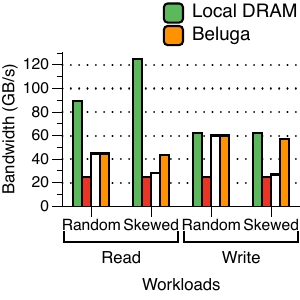}\label{fig:16kbw}}
  \hfill
  \subfloat[Median Latency (16K)]{\includegraphics[width=0.333\linewidth]{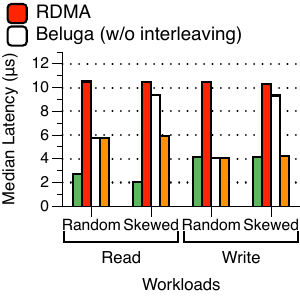}\label{fig:16kp50}}
   \hfill
  \subfloat[P99 Latency (16K)]{\includegraphics[width=0.333\linewidth]{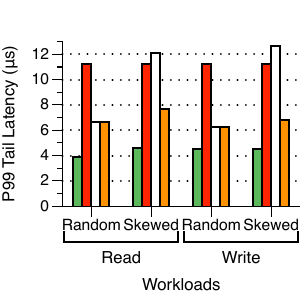}\label{fig:16kp99}}
  \mycaption{fig:16k}{\revisionx{Performance under concurrent 64B/16KB accesses (\textbf{Exp \#3})}}{\mdseries \revisionc{The setup consists of 64GB memory space with 16 threads (one thread per CPU core) performing synchronized memory access. Memory addresses are selected using zipf distribution, where the highly skewed workloads use parameter 0.99. }}
  \vspace{-0.1cm}
\end{figure}

\subsection{Performance under Complex Workloads}

\label{subsec:extra-bench}

\revisionx{
To evaluate \sys performance in complex scenarios, we conduct additional benchmarks focusing on concurrent and conflicting memory access patterns.
}

\noindent
\revisionx{
\textbf{(\textbf{Exp \#3}) Performance under skewed memory access.} We measure bandwidth, median latency, and p99 tail latency during concurrent memory access, and the results and configurations are shown in \autoref{fig:16k}. For local DRAM and CXL access, we use ntstore for write operations and clflush before read operations. 
This experiment also include the performance with 2~MB interleaving size and without memory interleaving, which demonstrates the necessity of enabling memory interleaving in skewed workloads. 
We have the following two observations. 
First, CXL demonstrates superior performance over RDMA. For median latency, CXL shows only 10.2\%\textasciitilde13.3\% of RDMA latency in 64B operations, and 39.5\%\textasciitilde56.2\% in 16KB operations. Notably, CXL achieves comparable write latency to local DRAM in 16KB write operation.
Second, CXL without memory interleaving exhibits lower bandwidth and higher latency under 16KB skewed workloads. This occurs because the first memory device in the CXL memory box becomes a bottleneck, introducing queuing latency during concurrent access.
}
% CXL performs better compared to RDMA, for median latency, compared to RDMA, the CXL is only 10.2\%\textasciitilde13.3\% in 64B, and 39.5\%\textasciitilde56.2\% in 16KB. And for 16KB write, CXL median latency is similar to local DRAM. 

\begin{figure}[t]
  \centering
  \subfloat[Performance with background write]{\includegraphics[width=0.5\linewidth]{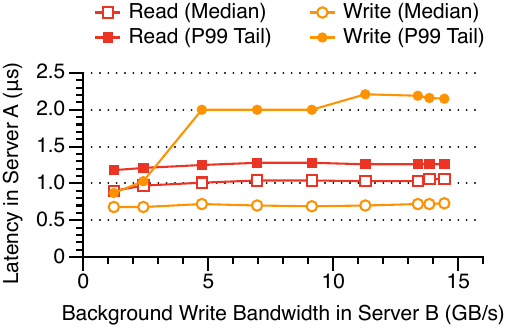}\label{fig:w-ab}}
  \hfill
  \subfloat[Performance with background read]{\includegraphics[width=0.5\linewidth]{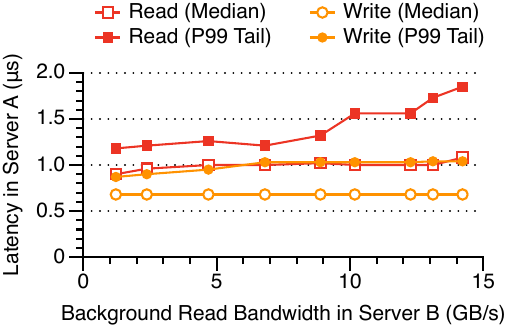}\label{fig:r-ab}}
   \hfill
  \mycaption{fig:16inter}{\revisionx{Performance with concurrent background workloads (\textbf{Exp \#4})}}{\revisionc{\mdseries It measures the 64B read/write latency while varying the background bandwidth from 0 to 15~GB/s on the same device. The RDMA latencies in (b) and (c) exceeded 8$\mu s$ and are omitted for better visualization.}}
  \vspace{-0.2cm}
\end{figure}

\noindent
\revisionx{
\textbf{(\textbf{Exp \#4}) Impact of background workloads.} The experiment applies fixed bandwidth pressure from a background server (server B) to a single memory device, while another server (server A) performs 64B read/write operations on the same device. \autoref{fig:16inter} illustrates read/write latency under varying background bandwidth pressure. Results show that while median latency remains stable regardless of background bandwidth, p99 latency increases when background bandwidth pressure operates in the same direction. This also demonstrates the bidirectional capability of CXL memory pools.
}

\section{Managing LLM KVCache with \sys}
\label{sec:kvcache}
In this section, we elaborate how to integrate \sys with a popular LLM inference framework (e.g., vLLM) and efficiently manage the corresponding KVCache.
As shown in Figure~\ref{pic:overview}, an LLM inference system typically consists of three key components: a large, shared memory pool to store the KVCache; a global index to map token blocks to their physical addresses in the pool; and a centralized scheduler to dispatch requests to different LLM instances. 

\sys-KVCache integrates with each of these components.
First, it introduces \sys into the KVCache management component, where the direct memory access interface provided by CXL greatly simplifies the KVCache access process (\S\ref{subsec:kvcache}).
Second, \sys-KVCache replaces the network communication between LLM instances and the indexing service via a CXL-based RPC (\S\ref{subsec:rpc}).
Lastly, with the flatter memory hierarchy in \sys-KVCache accelerates, the scheduler no longer needs to manage KVCache locality and focuses on computing resource allocation (\S\ref{subsec:scheduling}).

%First, by leveraging a direct non-contiguous access interface, \sys simplifies the shared memory pool access process (\S\ref{subsec:kvcache}). 
% Its global, unified address space with support for non-contiguous access significantly streamlines the logic required for KVCache store and load. 
%Second, \sys can improve the global index mechanism via CXL-based RPC. Given that current LLM frameworks use diverse prefix-indexing methods, we use \sys to simplify and accelerate communication between LLM instances and the index service, rather than attempting to design a new index in disaggregated CXL memory (\S\ref{subsec:rpc}). 
%Finally, by enabling a flatter memory hierarchy in \sys, the scheduler no longer needs to manage KVCache locality and can focus entirely on compute resource allocation (\S\ref{subsec:scheduling}).

% 方案22222222222222 END

% As our guidelines suggest, maximizing the benefits for each component may require applying different, workload-specific optimizations within \sys.

% \begin{enumerate}[leftmargin=*,label=(\arabic*)]
%     \item A large, \textbf{shared memory pool} on the CXL device for storing the raw KVCache blocks.
%     \item A lightweight, \textbf{distributed hash index} whose design is simplified by CXL's global address space. It directly maps request IDs to physical memory addresses, allowing it to be fully replicated on each node for fast, local lookups.
% \end{enumerate}

\begin{figure}[t]
  \begin{center}
    \includegraphics[width=0.9\linewidth]{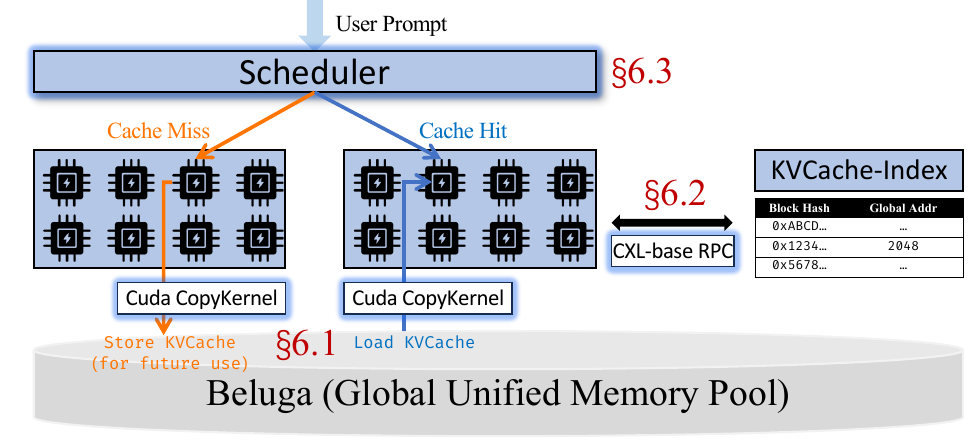}
  \end{center}
  \mycaption{pic:overview}{KVCache management with \sys}{} 
\end{figure}

% \subsection{A Lightweight KVCache Design}
% \subsection{Implementation}

% Figure~\ref{pic:overview} shows the typical architecture of LLM inference systems with KVCache offloading, and there are three components in it: a large, shared memory pool to store the KVCache, a global index to map the hash of token block to the addresses in shared memory pool, and a centralized scheduler to dispatch requests to different LLM inference instances. Based on the above design guideline, we observe all these components can be simplified or accelerated by \sys.

% First, for the shared memory pool, \sys provides global unified memory pool and also support non-contiguous memory access for storing and loading KVCache, therefore \sys can simply the KVCache access logic. Second, for the global index, we observe current LLM inference frameworks use different prefix-index to index the KVCache, therefore direct implement the index in CXL memory is not a general solution, therefore we use \sys to accelerate and simply the communication between LLM inference instances and the index service. 
% These two use cases may use different optimizations in \sys to maximize the performance benifits. 
% Third, due CXL memory is not a slow, tertiary tier, but a high-performance extension of local DRAM, the scheduler do not need to consider the locality of KVCache and focus on schedluing computing resouces.

\label{subsec:guildline}

\noindent
\subsection{Data Transfer Operations for KVCache}
%\subsection{KVCache Transfers in \sys-KVCache.}
\label{subsec:kvcache}

In transformer-based LLM inference systems, the KVCache for each token is typically stored in a highly fragmented memory layout.
As shown in \autoref{pic:layout}, a token's key and value tensors are typically stored non-contiguously across different attention layers. Furthermore, within a single layer, the KV tensors from different tokens are also stored non-contiguously.
However, an LLM inference system usually has to serialize the KVCache data (i.e. the key and value tensors), into contiguous blocks in a memory pool for effective use. 
This results in a significant KVCache transfer overhead.

KVCache transferring has two access patterns. 
\textit{Gather write (KVCache write)}: Data from many non-contiguous GPU locations is gathered into one contiguous block in the remote pool.
\textit{Scatter read (KVCache read)}: Data from one contiguous block in the remote pool is scattered to many non-contiguous GPU locations.

RDMA relies on scatter-gather lists (sglists) to handle the above transfers. 
Sglists allow combining multiple GPU memory chunks into a single RDMA request. However, for KVCache, sglists are inefficient.
For a Qwen-32B model with GQA (i.e., $n_{\text{heads}}\text{=}8$), a single token's KVCache may split into 128 non-contiguous chunks  (64 layers x 2 for key/value). %times
However, the hardware constraints (e.g., ConnectX-7 NIC limits sglists to 30 entries), resulting in complex logic to split the operation into multiple RDMA requests. 

Further, the data transfer problem becomes even more challenging when we exploit KVCache sparsity. Recent studies~\cite{DBLP:conf/nips/Zhang00CZC0TRBW23,DBLP:conf/nips/LiHYVLYCLC24} show that not all tokens contribute equally to generation. This shifts the "Scatter Read" access pattern into a more difficult non-contiguous-to-non-contiguous read.
For a Qwen-32B model with GQA($n_{\text{heads}}\text{=}8$), exploiting sparsity  results in 1024 small, 160-byte chunks for a single token: $n_{\text{chunks}} \text{=} n_{\text{layers}} \times n_{\text{heads}} \times 2$, where $n_{\text{layers}}\text{=}64, n_{\text{heads}}\text{=}8$. Accordingly, thousands of small requests are issued, which may trigger IOPS bottlenecks. In this case, RDMA is also inefficient.

In contrast, 
% \sys leverages CXL memory expansion to streamline KVCache movement, which brings two benefits.
% %
% First, \sys offers a unified address space. CXL memory is directly mapped into the host's virtual address space. 
% Second, 
\sys leverages the fine-grained custom copy kernel to handle unlimited gather writes (from multiple GPU regions to single CXL block) or scatter reads (from single CXL block to multiple GPU regions), eliminating the tedious request management in RDMA. 
% Second, \sys supports fine-grained sparse access. A single CUDA operation issues thousands of non-contiguous memory accesses to directly populate scattered KVCache chunks from CXL to GPU memory in the systems using KVCache sparsity.
%
Therefore, by shifting KVCache transfer duty from RDMA to CXL memory pool, \sys fundamentally resolves the KVCache migration bottleneck in an LLM inference system.

\noindent
\subsection{CXL-Based RPC}% in \sys-KVCache.}
% \noindent
% \textbf{}
\label{subsec:rpc}

For LLM inference systems with shared KVCache storage, efficient metadata management is critical. Most systems rely on centralized metadata services to track KVCache block locations, incurring frequent remote procedure calls (RPCs). These RPCs are conventionally realized by RDMA or TCP/IP protocols that incur high latency due to network stack overhead or kernel-user space transitions.

Different from the conventional RPC realization, \sys realizes the RPC via shared memory, fully utilizing the load/store semantics offered by CXL.
We reserve a small memory space in \sys to achieve cross-server communication.
\sys establishes the communication between a metadata server and clients in a producer-consumer manner. Initially, clients and the server pre-allocate fixed-size memory slots in CXL memory pool for request and response buffering. To produce a request, the client(i.e., producer) writes a request to an idle slot and updates a status flag (i.e., REQ\_READY). To consume a request, the server (i.e., consumer) continuously polls status flags using its internal spin-wait loops. After retrieving and processing a request, the server writes results to a dedicated reply slot and sets a RESP\_READY flag. The client obtains the response upon detecting the flag update. 

We also apply the following four optimizations to achieve high-performance.
First, non-temporal store (\texttt{ntstore}) instruction is adopted by the clients, avoiding CPU cache pollution. Second, the metadata server executes a \texttt{CLFLUSH} command before reading to ensure the visibility of the latest client-written data. 
\revisionc{Third, we also batch mfence operations for concurrent RPC requests, and ensure cache line alignment for the data in RPC. }
Further, during the whole producing-consuming process, \sys eliminates kernel-mode transitions and context switches by keeping all operations in user space. Putting these techniques together, \sys is able to achieve more efficient RPCs compared with RDMA or TCP/IP.
\begin{figure}[t]
  \begin{center}
    \includegraphics[width=0.95\linewidth]{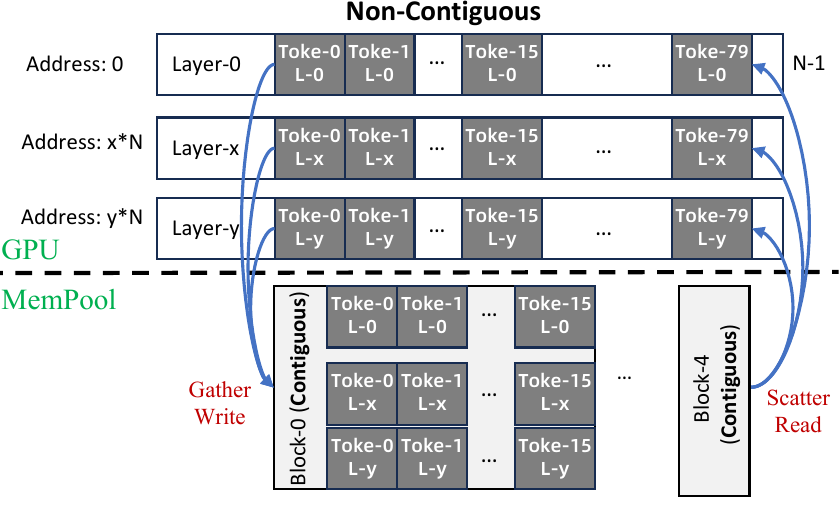}
  \end{center}
  \vspace{-0.25cm}
  \mycaption{pic:layout}{KVCache Layout in GPU and Memory Pool}{\revisionc{\mdseries A single KVCache block (16 tokens) in Qwen-32B (GQA) requires 128 non-contiguous 20KB data transfers.}} 
  \vspace{-0.15cm}
\end{figure}

\noindent
\subsection{Scheduling without KVCache Hierarchy} 
\label{subsec:scheduling}

%Traditional LLM inference systems relying on RDMA-based memory pools (e.g., NVIDIA's Dynamo~\cite{Dynamo} and Mooncake~\cite{mooncake,mooncake}) face inherent scheduling challenges, due to performance discrepancy between local and remote memory access. Accordingly, these systems have to include some well-designed cache-aware scheduling policies to handle the alleviate the performance discrepancy.  These policies usually designed to satisfy the KVCache locality constraints that require the requests had better routed to nodes hosting the needed KVCache blocks to avoid expensive remote memory fetches. In practice, these policies lead to some obvious drawbacks, including the scheduling complexity, load imbalance, and maintenance overhead~\cite{zuo2025cloudmatrix}. For example, such a policy may lead to skewed KVCache distribution, and the corresponding routers are difficult to recover from the skewed KVCache distribution.

Traditional LLM inference systems using RDMA-based memory pools, such as NVIDIA's Dynamo~\cite{Dynamo} and MoonCake~\cite{mooncake,mooncake}, suffer from a significant performance gap between local and remote memory access. To mitigate this gap, these systems must employ well-designed, cache-aware scheduling policies. 
These policies enforce KVCache locality by routing requests to nodes that already host the required data blocks, thereby avoiding expensive remote fetches. 
In practice, these policies lead to some obvious drawbacks, including the scheduling complexity, load imbalance, and maintenance overhead~\cite{zuo2025cloudmatrix}. 
For example, such a policy may lead to skewed KVCache distribution, and the corresponding routers are difficult to recover from the skewed KVCache distribution.

\sys largely eliminates the above constraints by leveraging CXL memory pools with near-local access latency. The entire CXL memory pool is abstracted as a unified and symmetric address space, where remote CXL memory access latency is comparable to local buffer access latency as demonstrated in \S\ref{subsec:basic-perfor}, eliminating the performance discrepancy incurred by RDMA. 
\revisionx{We} compared the performance of offloading KVCache to local memory versus \sys. The results show that for cache-hit requests, the TTFT of \sys is highly competitive with that of local memory.
% The experiment was conducted using an H20 GPU running the Qwen-32B model with the LV-EVAL workload. 
%The results show that for cache-hit requests, the TTFT when using local memory is approximately 20\% lower than when using \sys.
Therefore, the design in \sys enables cache-oblivious scheduling. 
% Such a scheduling operates independently of KVCache placement.
% as data access patterns do not affect performance; 
The incoming requests can be easily distributed using standard load-balancing techniques, ignoring cache locality.
Furthermore, LLM inference nodes can be added/removed without re-balancing KVCache partitions, as access overhead remains uniform.

By decoupling compute scheduling from KVCache locality, \sys significantly reduces scheduling overhead, and is able to maintain high throughput under skewed workloads. This paradigm shift enables LLM inference systems to prioritize computational efficiency over memory hierarchy optimization.

\section{Evaluation}
\label{sec:evaluation}

In this section, we seek to answer the following questions:

\begin{itemize}
[itemsep=0pt, parsep=0pt, labelsep=3pt, leftmargin=*, topsep=0pt, partopsep=0pt]
  % [leftmargin=*,itemsep=3pt]
% \begin{itemize}
\item \textbf{Overall performance:} How does \sys improve the end-to-end performance of LLM inference compared with a state-of-the-art RDMA-based system? (\S\ref{subsec:e2e})
\item \textbf{Sensitivity analysis:} How do workload characteristics and software/hardware configurations affect the performance benefits of \sys-KVCache? (\S\ref{subsec:sa})
% \item \textbf{Comparison with Local Memory:} What the performance gap between CXL memory pool and local memory? 
\item \textbf{Performance breakdown:} How does each component affect \sys-KVCache's overall performance? (\S\ref{subsec:otherperf})
\end{itemize}

\noindent
\textbf{Experimental setup.}
The evaluation is conducted on a cluster of two servers, each equipped with eight H20 \revisionb{(96~GB)} GPUs, for a total of 16 concurrent vLLM instances managed by a centralized scheduler. 
\revisionb{The vLLM version is V1 (v0.8.5), and we enable its prefix caching feature and leverage GPU HBM to store KVCache.} 
Other hardware configurations are shown in \autoref{tbl:env}. 
To evaluate the end-to-end impact of our design, we compare our \sys-KVCache against MoonCake (v3.2)~\cite{ mooncake, mooncake}, a highly-optimized RDMA-based baseline built upon the vLLM and LMCache framework (v0.3.1)~\cite{cachegen,vllm}, and \revisionx{Dynamo (v0.4.1), a RDMA-based baseline from Nvidia.}\footnote{Our evaluation is based on the versions available at the time of paper submission.}
\revisionx{For fairness, we limit the cache capacity in memory pool to 2~TB for both RDMA-based solutions and \sys.}

\noindent
\textbf{Model and workloads.}
We used the unquantized Qwen-32B model by default. Our primary workload was \revisionx{\textbf{LV-Eval}~\cite{yuan2024lv}}, which features long-context QA traces with all input sequences exceeding 15K tokens. To analyze the sensitivity of our system to context length, we created three variants of the LV-Eval workload by limiting the input context to 2K, 4K, and 8K tokens, named LV-Eval-2K, LV-Eval-4K, and LV-Eval-8K, respectively. We evaluate three metrics in these workloads: Time-to-First-Token (TTFT), Time-Per-Output-Token (TPOT), and Queries-Per-Second (QPS).

\subsection{End-to-End Performance Evaluation}
% \noindent
% \textbf{Methodology.}
\label{subsec:e2e}
To isolate the performance impact of initial cache population versus subsequent cache reuse, we designed two distinct experimental scenarios:
\begin{itemize}
[itemsep=0pt, parsep=0pt, labelsep=3pt, leftmargin=*, topsep=0pt, partopsep=0pt]
\item \textit{Cache-populate (first run):} This scenario measures the performance during the initial cache population, as vLLM computes and stores the KVCache into \sys. 
\item \textit{Cache-hit (second run):} All KVCache is pre-populated in the shared memory pool. This scenario measures the performance when the prefill stage is accelerated entirely by cache reuse.
\end{itemize}

\begin{table}[htbp]
\vspace{-0.14cm}
  \tblcaption{tb:vllm1}{\revisionx{Inference performance on LV-Eval~\cite{yuan2024lv} (\textbf{Exp \#5})}}{}
  \resizebox{0.8\linewidth}{!}{
  \begin{tabular}{l|c|c|c|c}
    \toprule
    \textbf{Metric} & \textbf{Dynamo} &  \textbf{vLLM}&\multicolumn{1}{c|}{\bf vLLM }   & \textbf{vLLM}\\
     &  & & \textbf{+MoonCake} & \textbf{+\sys}  \\
    \midrule
    \midrule
    \multicolumn{5}{c}{\bf First Run \revisionx{(Cache-populate)}}    \\
    \midrule
    Avg TTFT  & 17.96~s & 18.76~s & 19.66~s  & 17.22s  \\
    P99 TTFT    & 54.53~s  & 40.47~s & 41.65~s   & 44.6~s  \\
    \midrule
    Average TPOT   &1.55~s &  2.580~s & 1.97~s    & 1.54~s    \\
    P99 TPOT      &10.99~s  &  20.17~s & 20.84~s  & 16.14~s  \\
    \midrule
    QPS (req/s)  & 1.15  &  0.96 &  1.02   &   1.24  \\
    \midrule
    \midrule
    \multicolumn{5}{c}{\bf Second Run \revisionx{(Cache-hit)}}    \\
    % \textbf{Metric} & \textbf{vLLM v1} & \textbf{LMCache+MoonCake} & \textbf{CXL} & $\Delta$\\
    \midrule
    Average TTFT   & 15.69~s &  18.23~s & 13.00~s & 1.36~s  \\
    P99 TTFT     & 40.97~s  &  39.25~s & 39.91~s  & 5.02~s \\
    \midrule
    Average TPOT   & 1.38~s &  2.82~s & 1.10~s    & 0.15~s     \\
    P99 TPOT      & 11.01~s &  19.74~s & 10.58~s  & 1.34~s   \\
    \midrule
    QPS (req/s)  & 1.32  &  0.96 &  1.54  &  11.32  \\
    \bottomrule
  \end{tabular}
  \vspace{-0.16cm}
  }
\end{table}

\noindent
\textbf{(\textbf{Exp \#5}) Results.} The evaluation uses a closed-loop client model to measure the peak throughput. 
As shown in \autoref{tb:vllm1}, \sys-KVCache consistently outperforms the RDMA-based baseline across all metrics.
In the cache-populate scenario, where the workload has a 30\% cache hit ratio, both MoonCake and \sys-KVCache outperform the original vLLM. Notably, \sys-KVCache further improves upon MoonCake, reducing the average TTFT by 12.4\% and increasing QPS by 21.5\%.
% This demonstrates that even during the costly process of populating the cache, our CXL-based approach is more efficient.
The performance advantage is most significant in the cache-hit scenario. \sys-KVCache reduces the average TTFT by 89.6\% and achieves a 7.35$\times$ improvement in QPS. 
MoonCake's lower performance can be attributed to two main factors. First, there exists an inherent performance gap between CXL and RDMA technologies. \revisionb{Second, its implementation includes additional overhead in the critical path, such as extra memory copies and allocations, which are being continuously addressed by the community~\cite{mooncake}. These issues highlight the programming complexity introduced by RDMA-based memory pools.}

% These results confirm that the superior end-to-end performance of \sys stems from its efficiency in both the initial cache population and subsequent cache reuse.

\noindent
\revisionb{
\textbf{Cache Hit Ratio in GPU}. In our experimental setup, the model occupies 60.0~GB, we allocate 92\% of the total memory space, therefore there are 28.3~GB for KVCache storage. Our performance analysis shows that the cache hit ratio in GPU HBM varies dynamically throughout the execution, reaching a peak performance of 14.6\% under this memory configuration.
}

\subsection{Sensitivity Analysis}

\label{subsec:sensitivity}
In this section we specifically analyze the sensitivity of our systems across three critical dimensions: 
the workload characteristics, software configurations, and hardware configurations.

\noindent
\textbf{Performance under different workload characteristics.} 
\revisiona{To assess \sys performance under realistic scenarios, we design two test workloads that vary in request arrival rates and input lengths.}

\noindent
\underline{(\textbf{Exp \#6}) Request arrival rates.}  
Our evaluation methodology consists of two phases. First, we execute the LV-EVAL workload to initialize the KVCache. Then, we conduct performance measurements with request rates ranging from 0.3 to 9.0 QPS during a second execution. 
\autoref{pic:latvstp} shows the results of TTFT and TPOT. We observe that \sys-KVCache consistently outperforms MoonCake, achieving lower latencies for both metrics. This is because, on the second run, all requests result in a cache hit, making the KVCache read time the primary system bottleneck. In this cache-bound scenario, the CXL-based data access in \sys is significantly more efficient than the RDMA-based access used by MoonCake.

\begin{figure}[t]
  \centering
  \subfloat[TTFT with different Request Rates]{\includegraphics[width=0.5\linewidth]{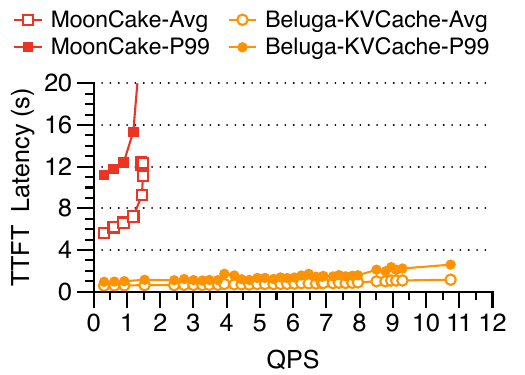}\label{fig:ttft}}
  \hfill
  \subfloat[TPOT with different Request Rates]{\includegraphics[width=0.5\linewidth]{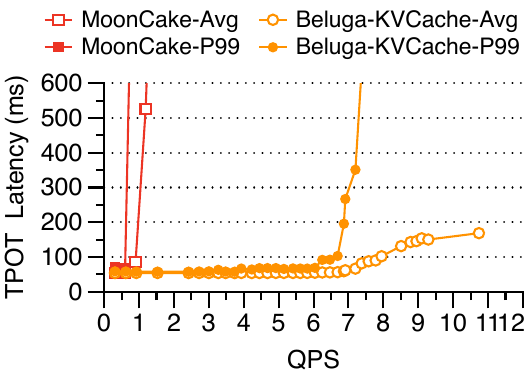}\label{fig:tpot}}
  \vspace{-0.15cm}
  \mycaption{pic:latvstp}{Sensitivity to request arrival rates (\textbf{Exp \#6})}{}
   \vspace{-0.2cm}
\end{figure}
% \vspace{-0.3cm}
\begin{figure}[t]
  \centering
  \subfloat[Average TTFT]{\includegraphics[width=0.48\linewidth]{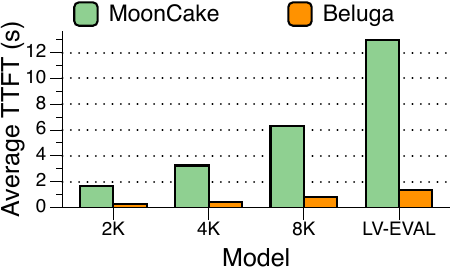}\label{fig:avg}}
  \hfill
  \subfloat[P99 TTFT]{\includegraphics[width=0.48\linewidth]{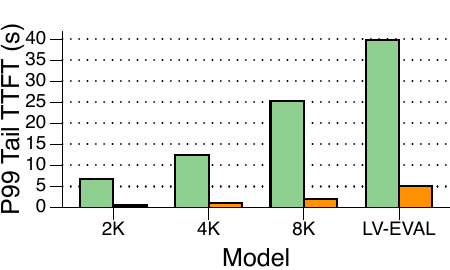}\label{fig:p99}}
   \vspace{-0.15cm}
  \mycaption{pic:length}{Sensitivity to input context lengths (\textbf{Exp \#7})}{}
   \vspace{-0.5cm}
\end{figure}

\begin{figure}[t]
  \centering
  \subfloat[P-D Ratio (QPS)]{\includegraphics[width=0.333\linewidth]{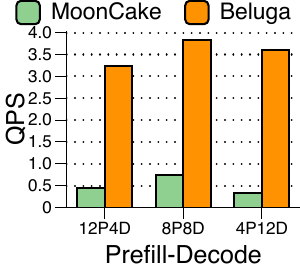}\label{fig:pda}}
  \hfill
  \subfloat[P-D Ratio (TTFT)]{\includegraphics[width=0.333\linewidth]{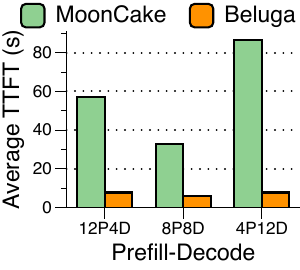}\label{fig:pdb}}
  \hfill
  \subfloat[Block sizes (TTFT)]{\includegraphics[width=0.333\linewidth]{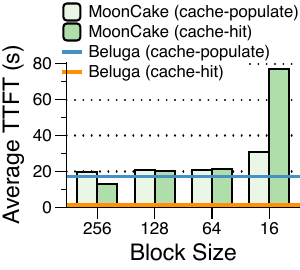}\label{fig:block}}
  \vspace{-0.15cm}
  \mycaption{pic:mix}{Sensitivity to software configurations (\textbf{Exp \#8})}{}
  \vspace{-0.2cm}
\end{figure}

\noindent
\underline{(\textbf{Exp \#7}) Input context lengths.}
We evaluate the systems using workloads with varying input context lengths: 2K, 4K, and 8K, via limiting the input length of the original LV-EVAL. \autoref{pic:length} shows the average and P99 TTFT for both \sys-KVCache and MoonCake.
We observe that as the number of input tokens increases, the performance improvement of \sys becomes more significant. This is because the KVCache write/read time accounts for a larger portion of the end-to-end latency in longer-context scenarios.

\noindent
\revisionx{
\textbf{(\textbf{Exp \#8}) Performance with different software configurations.}
We evaluate \sys performance under different inference framework configurations, focusing on two aspects: Prefill-Decode deployment and KVCache block size configurations.}

\noindent
\revisionb{
\underline{Prefill-decode disaggregated.}
In the prefill-decode disaggregated architecture, KVCache is first stored in the memory pool after the prefill phase, then loaded by decode nodes. As demonstrated in Fig.\ref{fig:pda} and Fig.\ref{fig:pdb}, \sys's optimized KVCache load/store path achieves 3.41$\times$\textasciitilde9.47$\times$ higher QPS compared to MoonCake.
}

\noindent
\revisionb{
\underline{Block size of KVCache access.} 
Our experiments reveal significant differences in block size requirements between RDMA-based solutions and \sys. RDMA-based approaches require operation batching to amortize control overhead, combining multiple vLLM KVCache blocks into larger super blocks. While vLLM uses a 16-token block size for efficient GPU HBM space management, this granularity is inefficient for RDMA transfers. Consequently, LMCache defaults to a larger block size of 256 tokens for KVCache indexing and transfer.
As shown in Fig.\ref{fig:block}, MoonCake achieves 13.0s TTFT for cache hits with the larger block size (256 tokens). However, when using the smaller 16-token block size, TTFT increases to 76.8s, exceeding even the recomputation latency of the first run. In contrast, \sys operates efficiently with vLLM's native block size, eliminating the need for operation batching.    
}

\noindent
\revisionx{
\textbf{Performance with and without memory interleaving.} 
We conduct an experiment to demonstrate the effectiveness of software memory interleaving in \sys. With software interleaving across two CXL/PCIe adapters and 32 memory devices, the system achieves a QPS of 11.32 during rerun (cache hit), a 33.2\% improvement compared to 8.49 without interleaving. The results confirm that software interleaving effectively distributes memory access load and reduces access contention, leading to better overall system performance.
}

% \subsection{Comparison with Local Memory}
% In this section, we analyze the performance of \sys compared with the local memory. Due to the local memory can not be direct shared by the remote server. In this section we only use 8 GPUs in a single server. 
% We implement a DRAM-KVCache, which replace the resource in CXL memory pool with the local DRAM. The index and all 8 LLM instances are located in the same server. This experiment shows the performance gap between local DRAM and the CXL memory pool. The workloads use LV-Eval, LV-Eval-4K, LV-Eval-8K. 

\label{subsec:sa}
\subsection{The Breakdown of Performance Improvement}
% \subsection{Analysis of Performance Gains}
% \subsection{Contribution of Each Component}
\label{subsec:otherperf}

\revisiona{In this section we analyze the breakdown of performance improvement of our systems, including KVCache transfer performance and RPC performance.}

\begin{figure}[t]
\centering
\subfloat[Write Latency]{\includegraphics[width=0.49\linewidth]{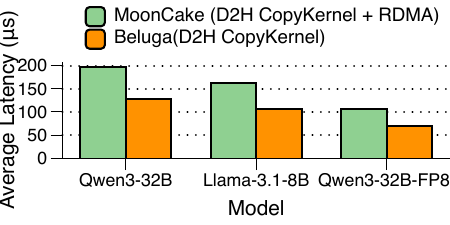}\label{fig:CW}}
\hfill
\subfloat[Read Latency]{\includegraphics[width=0.49\linewidth]{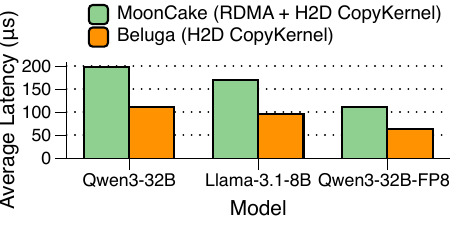}\label{fig:CR}}
% \hfill
% \subfloat[Read]{\includegraphics[width=0.33\linewidth]{fig/ComplexX.pdf}\label{fig:CX}}
\vspace{-0.15cm}
\mycaption{pic:complex}{Data transfers for dense KVCache (\textbf{Exp \#9})}{}
\vspace{-0.15cm}
\end{figure}

\noindent
\textbf{(\textbf{Exp \#9}) Performance of data transfers for KVCache (Dense).} 
\revisiona{To demonstrate the advantages of CXL over RDMA in KVCache transfers, we conducte benchmarks focusing on scatter-gather operations - the crucial operations in dense KVCache transfers.}
Our evaluation encompasses three different models: Qwen3-32B, Llama-3.1-8B, and Qwen3-32B-FP8. Each model exhibits distinct KVCache block layouts. For instance, in the Qwen-32B model (using Grouped Query Attention, or GQA), a KVCache block is distributed across 128 non-contiguous sub-blocks, while Llama-3.1-8B utilizes 64 sub-blocks.
Across all models, each KVCache block maintains 16 tokens, which is vLLM's default configuration. This token count remains constant and does not influence the number of non-contiguous sub-blocks in the models.
As shown in \autoref{pic:complex}, compared to MoonCake, which uses a CPU-centric method with an inefficient two-step data path (GPU → Host → MemPool), \sys eliminates the host memory bounce buffer entirely. This direct data path reduces KVCache write and read latencies by 36.2\% and 38.7\%, respectively, showcasing its superior efficiency and simpler data management.
% Even when compared to a more optimized GPUDirect RDMA approach, which avoids the host buffer, \sys still delivers a 5.7\% reduction in scatter-read (load) latency and a 5.3\% reduction in gather-write (store) latency, 

\noindent
\textbf{(\textbf{Exp \#10}) Data transfers for KVCache (Sparse).}
\revisiona{To further demonstrate CXL's efficiency in KVCache transfer, we extend our evaluation to models with sparse KVCache implementations.}

\begin{table}[h]
    \centering
    \tblcaption{tb:topktokens}{Sparsity analysis and the read latency (\textbf{Exp \#10})}{}
    \resizebox{0.88\linewidth}{!}{
    \begin{tabular}{cc|r|r}
        \toprule 
         \multicolumn{2}{c|}{} & \textbf{Llama-3-8B} & \textbf{Qwen3-32B} \\
        \midrule
        \multirow{1}{*}{Top 256 tokens} & Non-contig & 131330 & 782774 \\
          (per head and per layer) & Contiguous & 130814 & 265802 \\
        \midrule
        \multirow{2}{*}{Loading 16 tokens (~$\mu s$)}  & RDMA & 2670 & 5260 \\
        & CXL & \textbf{97} & \textbf{211}  \\
        \bottomrule
        
    \end{tabular}
    }
    \vspace{-0.2cm}
\end{table}

We first analyze the KVCache sparsity patterns in both Qwen-32B and Llama-3-8B models using an attention score-based sparsification method~\cite{DBLP:conf/nips/Zhang00CZC0TRBW23}. This analysis helps us understand the unique characteristics of sparse workloads in large language models.
\autoref{tb:topktokens} shows the results of selecting the top 256 tokens per head and per layer for a sequence with 7942 tokens. Our analysis reveals that over 74\% of the selected high-importance tokens in Qwen-32B are non-contiguous, confirming that the memory access pattern for sparse KVCache is overwhelmingly discrete.

Based on these findings, we evaluate the performance of reading KVCache for 16 sparse tokens. In the Qwen-32B model, \sys achieves a remarkable 95.9\% latency reduction compared to RDMA. The reason is fundamental: the RDMA solution is bottlenecked by issuing numerous high-overhead requests for non-contiguous data, while \sys allows a single CUDA kernel to manage the entire fine-grained transfer efficiently through CXL's directly mapped memory.

\begin{figure}[b]
  \vspace{-0.3cm} 
  \centering
  \subfloat[Latency]{\includegraphics[width=0.5\linewidth]{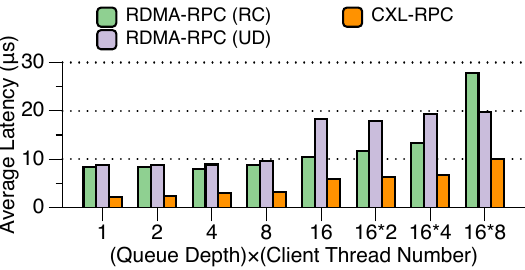}\label{fig:rpc-lat}}
  \hfill
  \subfloat[Throughput]{\includegraphics[width=0.5\linewidth]{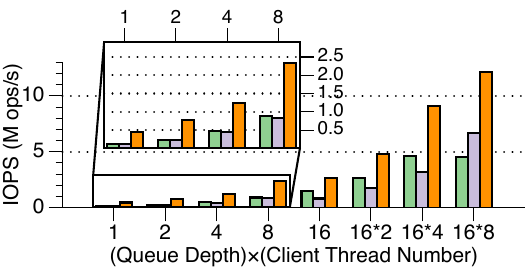}\label{fig:rpc-tp}}
  \vspace{-0.1cm} 
  \mycaption{pic:rpc}{Performance of CXL/RDMA RPC (\textbf{Exp \#11})}{}
  \vspace{-0.2cm}
\end{figure}

\noindent
\textbf{(\textbf{Exp \#11}) Performance of RPC.} 
\revisiona{In this evaluation, we measure how much faster RPC performs when using CXL instead of RDMA.}
We compare our CXL-RPC against two functionally equivalent RDMA implementations: RDMA-RC (Reliable Connection) and RDMA-UD (Unreliable Datagram). The RDMA baseline also uses busy-polling on its completion queue (CQ), just similar to the thread-model of our CXL-RPC.
We use a ping-pong benchmark between two servers, with a single-threaded RPC service and multi-threaded clients. Each RPC request and reply is 64 bytes. 

The results are presented in \autoref{pic:rpc}. At low concurrency (QD=1), CXL-RPC achieves a round-trip latency of 2.11~$\mu s$, \revisionb{which involves four memory operations (two reads and two writes) and queuing overhead}
It achieves a 4$\times$ improvement over both RDMA-RC (8.39~$\mu s$) and RDMA-UD (8.83~$\mu s$). 
This advantage stems from CXL's direct load-store access model, which bypasses the extensive software and hardware overhead inherent in RDMA protocols. 
Under high concurrency (QD=128), CXL-RPC's throughput (\revisionx{single thread}) reaches 12.13\,Mops, significantly outperforming both RDMA-RC (4.5\,Mops) and RDMA-UD (6.65\,Mops) by 2.7$\times$ and 1.8$\times$, respectively. This demonstrates how CXL leverages the CPU's cache hierarchy to buffer and burst writes, effectively amortizing protocol overhead under heavy load. 
\revisionb{It is important to note that CXL-RPC in \sys provides lower reliability guarantees compared to RDMA transport protocols. Our implementation relies on upper-layer mechanisms for reliability assurance. While strengthening these guarantees remains a topic for future research, the current CXL-RPC design in \sys prioritizes performance and cost efficiency within rack-scale deployments, rather than serving as a complete RDMA replacement.
}
\section{Future Work on CXL Switches}
\label{sec:future}

Using one of the first commercial CXL 2.0 switches, this paper presents a foundational analysis of a CXL-switched memory pool. This paper establishes key performance baselines for CXL 2.0 systems. It also demonstrates CXL's effectiveness for complex tasks like KVCache offloading. These results open up a vast design space for future CXL-native systems. 
Based on our analysis, we outline several promising directions for future research.

\noindent
\textbf{Future work on hardware architecture.}
The emergence of CXL 3.1 suggests a path toward a large-scale, fully disaggregated architecture. We envision a future GPU cluster built on a CXL fabric, as shown in \autoref{pic:planc}. In this architecture, a symmetric fabric would connect pools of compute, memory, and storage, providing unified, low-latency access for all attached devices. 
By connecting GPUs directly to this fabric, the latency overhead of traversing the host's PCIe switch and root complex is eliminated, providing lower-latency memory access compared to the architecture in \sys.
%Compared to the architecture used in \sys, this new design could offer GPUs lower-latency access to the memory pool.

% While this architecture is attractive, its practical implementation faces significant technical challenges. We identify two main areas for future research:

\noindent
\textbf{Future work on software design.}
This new hardware architecture requires a corresponding software stack to manage resource pooling and coherent data sharing at scale.
First, it needs to develop resource schedulers that can dynamically manage GPU/CPU resources across multiple CXL switches. This requires a co-design approach that balances application needs with the specific performance characteristics of the hardware.
Second, a critical research area is the design of more efficient coherence protocols for CXL memory pools. Potential directions include: (1) leveraging application-level semantics to relax coherency, (2) implementing scalable, directory-based coherence using in-switch resources, and (3) developing hybrid models where a small, hardware-coherent region manages coherence metadata for a significantly larger address space. 

\noindent
\revisionx{
\textbf{Future work on database design.} CXL memory pooling enables new opportunities for memory-intensive data management systems that require large-scale random access capabilities, particularly for vector databases and graph databases. For instance, graph-based algorithms like HNSW, which traditionally demand full in-memory storage, can leverage CXL memory pools to break through memory capacity limitations. 
}

\section{Related Work}
\label{sec:relatedwork}

 \begin{figure}[t]
   \begin{center}
     \includegraphics[width=0.8\linewidth]{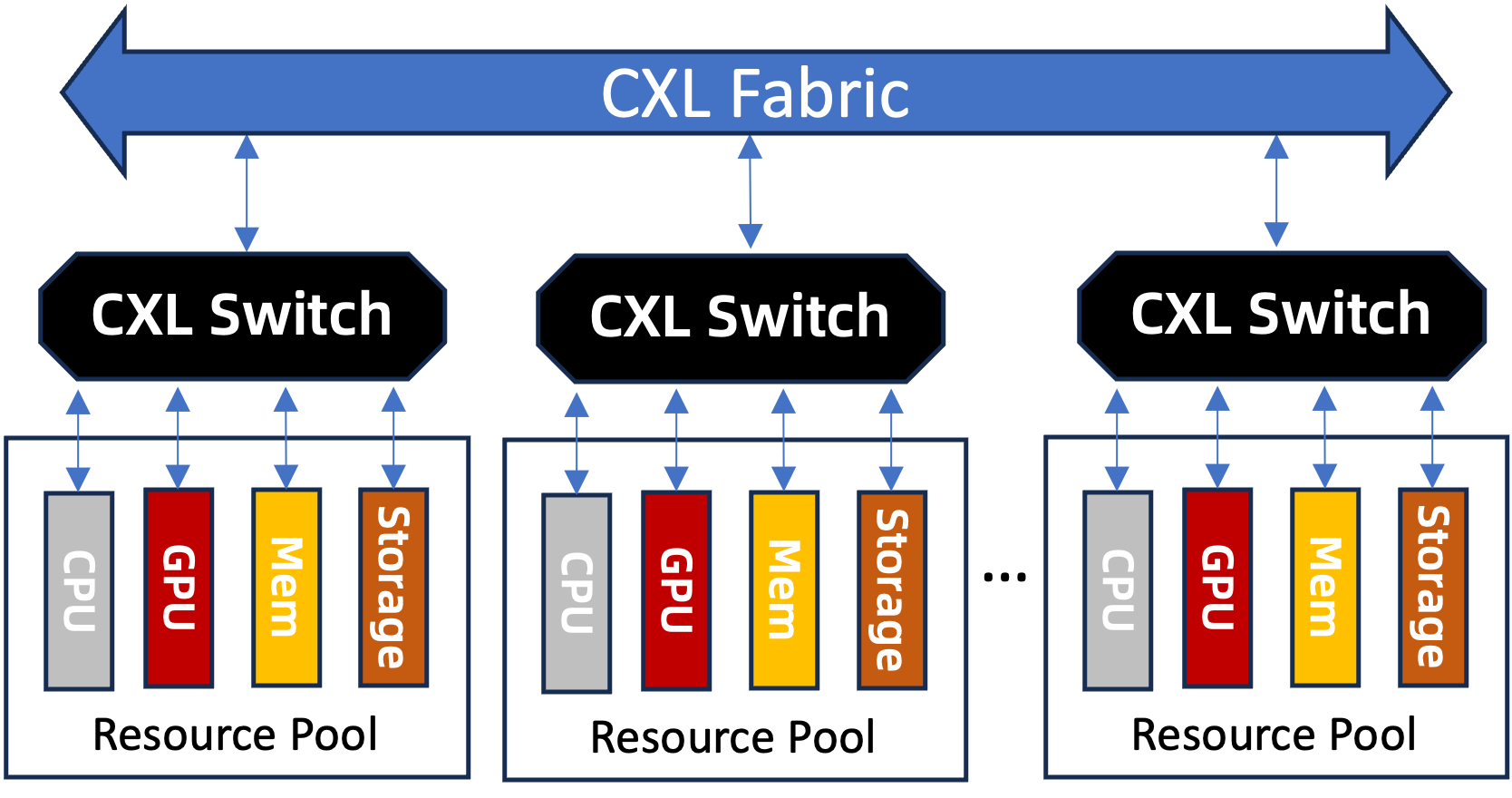}
    \end{center} % pic:perf1}{4MB I/O Latency
   \vspace{-0.2cm}
  \mycaption{pic:planc}{Fully disaggregated architecture}{} 
   \vspace{-0.1cm}
\end{figure}

\noindent
\textbf{Data management in LLM inference.}
KVCache represents the primary memory bottleneck in LLM inference~\cite{DBLP:journals/corr/abs-2407-18003,DBLP:conf/nips/LiHYVLYCLC24,DBLP:conf/iclr/Ge0LZ0024}, motivating extensive research efforts aimed at reducing its footprint. 
% At the algorithmic level, researchers have explored novel attention architectures~\cite{DBLP:journals/corr/abs-1911-02150,DBLP:conf/emnlp/AinslieLJZLS23}, cache compression strategies~\cite{DBLP:conf/nips/Zhang00CZC0TRBW23,DBLP:conf/iclr/Ge0LZ0024}, and quantization techniques~\cite{DBLP:conf/nips/HooperKMMSKG24}. MQA~\cite{DBLP:journals/corr/abs-1911-02150} and GQA~\cite{DBLP:conf/emnlp/AinslieLJZLS23} reduce the computation complexity of KVCache via architectural optimizations. H2O~\cite{DBLP:conf/nips/Zhang00CZC0TRBW23}, FastGen~\cite{DBLP:conf/iclr/Ge0LZ0024}, and SnapKV~\cite{DBLP:conf/nips/LiHYVLYCLC24} exploit the sparsity of attention scores to selectively compress or evict less important KVCache entries.
At the algorithmic level, 
researchers have explored various optimizations: MQA~\cite{DBLP:journals/corr/abs-1911-02150} and GQA~\cite{DBLP:conf/emnlp/AinslieLJZLS23} reduce KVCache computation complexity through architectural innovations, while H2O~\cite{DBLP:conf/nips/Zhang00CZC0TRBW23}, FastGen~\cite{DBLP:conf/iclr/Ge0LZ0024}, and SnapKV~\cite{DBLP:conf/nips/LiHYVLYCLC24} leverage attention score sparsity to selectively compress or evict less important KVCache entries.
At the system level, PagedAttention~\cite{vllm} introduced a virtual memory-inspired mechanism for efficient KVCache management, while MoonCake~\cite{mooncake,mooncake} and Dynamo~\cite{Dynamo} leveraged remote resource pooling to enable large-scale request scheduling.
Building upon these insights, \sys proposes a new hardware architecture for efficient KVCache management that reduces data movement overhead at the system level while enabling flexible random access to support the current novel algorithmic optimizations, such as the sparse KVCache.

% Beyond these efforts, hardware-level optimizations have become indispensable for addressing the memory demands of LLM inference.

% MQA~\cite{DBLP:journals/corr/abs-1911-02150} and GQA~\cite{DBLP:conf/emnlp/AinslieLJZLS23} reduce the computation complexity of KVCache via architectural optimizations. H2O~\cite{DBLP:conf/nips/Zhang00CZC0TRBW23}, FastGen~\cite{DBLP:conf/iclr/Ge0LZ0024}, and SnapKV~\cite{DBLP:conf/nips/LiHYVLYCLC24} exploit the sparsity of attention scores to selectively compress or evict less important KVCache entries.
% Algorithmic optimizations are insufficient to meet the ever-growing memory requirements of LLMs, shifting the research focus toward the underlying system layer~\cite{vllm,mooncake,Dynamo}. 

% \vspace{0.1cm}
\noindent
\textbf{Disaggregated memory architectures for data management. }
Disaggregated memory as a resource extension has been widely adopted in database systems. RDMA-based memory disaggregation~\cite{DBLP:conf/fast/Li0ZCS23,DBLP:journals/pvldb/CaiGZACOTTW18,DBLP:conf/osdi/YuJKKC22, bin1, bin2, bin3, bin4} has been applied in various scenarios, including transaction processing and storage optimization~\cite{DBLP:conf/fast/Li0ZCS23,DBLP:journals/pvldb/LuHLWL24,DBLP:conf/sosp/LuoSZ0LZ24}. Recent systems like MoonCake~\cite{mooncake,mooncake} and Dynamo~\cite{Dynamo} leverage RDMA to offload KVCache management to memory pools.
While RDMA provides one-sided memory access, it operates as a network protocol rather than a true memory interface, introducing additional complexity and latency overhead. CXL addresses these limitations by offering native memory-like interfaces and characteristics.
There also exist specialized solutions like CloudMatrix~\cite{zuo2025cloudmatrix}, a large-scale supernode built on Unified Bus (UB) in Huawei. While it establishes a CXL-like fabric for uniform peer-to-peer communication, its vendor-specific design limits widespread adoption.

% \vspace{0.1cm}
\noindent
\textbf{CXL-based memory architectures.}
The introduction of CXL has triggered extensive research into its performance characteristics and system-level potential. Early characterization studies~\cite{cxlrack, cxlpci, DBLP:conf/asplos/LiuHWBNJNL25, pond, gouk2022direct} have leveraged FPGA-based prototypes or initial CXL 1.1 hardware to provide the first quantitative analysis of this emerging interconnect. Subsequently, a range of studies have demonstrated the practicality of CXL across diverse real-world scenarios, encompassing in-memory databases~\cite{DBLP:conf/damon/AhnCLGKJRPMK22, polarcxl,DBLP:conf/icde/Guo024}, graph processing~\cite{DBLP:conf/sc/SanoBHKSNTKS23}, and deep learning systems~\cite{DBLP:conf/isca/LiuZHZLWLJX24,DBLP:conf/icpp/ArifARV22}, among others. Moreover, several works~\cite{DBLP:journals/corr/abs-2405-14209,cxlrack,DBLP:journals/corr/abs-2308-02501} have conducted careful comparative analyses of CXL and RDMA. For example, \cite{DBLP:journals/corr/abs-2405-14209} demonstrates the performance superiority of CXL over RDMA, and \cite{cxlrack} reports on leveraging CXL to enhance the performance of RDMA-based systems. 
Prior work has explored CXL for CPU memory expansion. This paper extends CXL's capabilities to GPU-centric KVCache management using commercial CXL 2.0 switches.

\section{Conclusion}
\label{sec:conclusion}

In this paper, we have designed, evaluated, and optimized Beluga, a novel CXL-switch-based memory system for GPU clusters. Through a detailed characterization of commercial CXL hardware, we identify key performance characteristics and propose a set of optimizations for building high-performance applications. To demonstrate the effectiveness of Beluga, we design and implement Beluga-KVCache, a system tailored for managing the large-scale KVCache in LLM inference. 
This system includes optimized and simplified data transfers, a lightweight RPC, and a simplified scheduler based on Beluga. 
% Experiments show that \sys reduces time-to-first-token by up to 82.7\% and achieves up to 4.79$\times$ higher throughput than a state-of-the-art RDMA baseline. These results demonstrate the potential of CXL as a foundation for future disaggregated memory systems.
Our evaluation demonstrates that Beluga-KVCache significantly outperforms the state-of-the-art RDMA-based solution in MoonCake, highlighting CXL's potential as a foundation for future disaggregated memory systems.

% \newpage

\section*{Acknowledgements}
We sincerely thank anonymous reviewers for their valuable feedback, which significantly improved this paper. We thank the PolarDB Infrastructure Team and Alibaba Infrastructure Service (AIS) Group for their support and contributions to this work.

%-------------------------------------------------------------------------------
\bibliographystyle{unsrt}
\bibliography{sample.bib}

% \appendix
% \appendix

%%%%%%%%%%%%%%%%%%%%%%%%%%%%%%%%%%%%%%%%%%%%%%%%%%%%%%%%%%%%%%%%%%%%%%%%%%%%%%%%

%%%%%%%%%%%%%%%%%%%%%%%%%%%%%%%%%%%%%%%%%%%%%%%%%%%%%%%%%%%%%%%%%%%%%%%%%%%%%%%%
\end{document}